\documentclass[conference]{IEEEtran}
\IEEEoverridecommandlockouts
\usepackage{cite}
\usepackage{amsmath,amssymb,amsfonts}
\usepackage{algorithmic}
\usepackage{subcaption}
\usepackage{graphicx}
\usepackage{url}
\usepackage{textcomp}
\usepackage{comment}
\usepackage{xcolor}
\usepackage{balance}
\usepackage{csquotes}
\newcommand{\etal}{{\em et al.}}
\newcommand{\peter}[1]{\textcolor{green}{#1}}

\newcommand{\ahmad}[1]{\textcolor{blue}{#1}}

\newcommand{\change}[1]{\textcolor{black}{#1}}
\def\BibTeX{{\rm B\kern-.05em{\sc i\kern-.025em b}\kern-.08em
    T\kern-.1667em\lower.7ex\hbox{E}\kern-.125emX}}
\begin{document}

\title{
The Seeker’s Dilemma: Realistic Formulation and Benchmarking for Hardware Trojan Detection
\author{Amin Sarihi$^1$, Ahmad Patooghy$^2$, Abdel-Hameed A. Badawy$^1$, and Peter Jamieson$^3$\\
    $^1$Klipsch School of Electrical and Computer Engineering, New Mexico State University, Las Cruces, NM\\
    $^2$Department of Computer Systems Technology, North Carolina A\&T State University, Greensboro, NC\\
    $^3$Department of Electrical and Computer Engineering, Miami University, Oxford, OH\\
    ~$^1$\{sarihi, badawy\}@nmsu.edu,~$^2$apatooghy@ncat.edu,~$^3$jamiespa@miamioh.edu}}



\maketitle
\thispagestyle{plain}
\pagestyle{plain}


\begin{abstract}
This work focuses on advancing security research in the hardware design space by formally defining the realistic problem of Hardware Trojan (HT) detection. The goal is to model HT detection more closely to the real world, \textit{i.e.}, describing the problem as \textquote{The Seeker’s Dilemma} (an extension of Hide\&Seek on a graph), where a detecting agent is unaware of whether circuits are infected by HTs or not. Using this theoretical problem formulation, we create a benchmark that consists of a mixture of HT-free and HT-infected restructured circuits while preserving their original functionalities. The restructured circuits are randomly infected by HTs, causing a situation where the defender is uncertain if a circuit is infected or not. We believe that our innovative dataset will help the community better judge the detection quality of different methods by comparing their success rates in circuit classification. We use our developed benchmark to evaluate three state-of-the-art HT detection tools to show baseline results for this approach. We use Principal Component Analysis to assess the strength of our benchmark, where we observe that some restructured HT-infected circuits are mapped closely to HT-free circuits, leading to significant label misclassification by detectors.
\end{abstract}


\section{Introduction}

Hardware Trojans (HTs) are a serious threat to digital electronics in general and the design and manufacturing of Integrated Circuits (ICs) in particular. An HT is an unwanted modification in the design or manufacturing of an IC such that the chip’s expected behavior is altered. Such modifications mostly follow malicious goals such as denial-of-service or information stealing~\cite{yu2021hw2vec}. 

Potential impacts of security breaches through HTs have pushed many researchers to look into the methods and algorithms to detect HTs in ICs in the early stages of the design and manufacturing process~\cite{shakya2017benchmarking}. The two main approaches are test-time (deploy time) HT detection (the detection of the existence of an HT on an IC) with several methods~\cite{jain2021survey} and design-time detection (the detection of the presence and location of an HT in a circuit netlist) with a range of methods~\cite{chakraborty2009mero, hasegawa2017trojan, pan2021automated, gohil2022deterrent,yu2021hw2vec,sarihi2023multi}. Although many HT detection methods have been proposed in the literature, the field still needs a formal definition of the problem that mirrors the real-world problem of HT detection.

\begin{figure*}[!ht]
    \centering
    
    \includegraphics[scale=0.51]{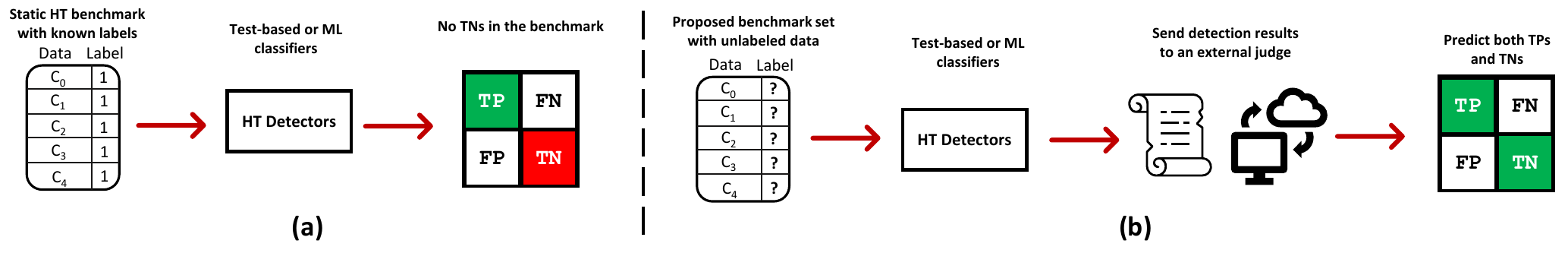}
    \caption{Comparison of a) current HT detection approaches with static HT benchmarks vs. b) our proposed HT detection flow, including \change{restructured} benchmarks with and without HTs}
    \label{fig:benchmark}
    \vspace{-6mm}
\end{figure*}

In this paper, our goal is to advance the state-of-the-art in HT detection by formally defining the problem of HT insertion/detection in digital ICs as \change{a game between two players}. Our comprehensive problem statement is rooted in the \textit{Hide\&Seek} problem on a graph. We call this new formulation “The Seeker’s Dilemma” as it more closely resembles the problem of HT detection from the perspective of real-world manufacturers and distributors of ICs. \change{We take “The Seeker’s Dilemma” approach in creating realistic HT benchmarks to address the shortcomings of existing HT benchmarks. The HT location and size are known to the security researcher in current benchmarks such as~\cite{trusthub,cruz2018automated,gohil2022attrition}. This enables the defense side to fine-tune their HT detectors to showcase strong HT detection rates. On the contrary, we believe that a standard benchmark should contain several HT-infected and HT-free instances to provide better dataset balance for researchers. Figure~\ref{fig:benchmark} shows the difference between existing benchmark approaches and our proposed method.}
Next, we define a generic evaluation methodology for HT detection with “The Seeker’s Dilemma” to help the community fairly evaluate existing and future HT detection methods. To facilitate this, we provide a new benchmark and use it to evaluate three state-of-the-art detectors: HW2VEC~\cite{yu2021hw2vec} trained with two different training sets, RL\_HT\_DETECT~\cite{sarihi2023trojanframework}, and DETERRENT~\cite{gohil2022deterrent}. Our released HT benchmark, called \textit{Seeker1}, is available at~\cite{githubGitHubNMSUPEARLSeekersDilemmaHardwareTrojanBenchmarks} so researchers can test their tools. 

We anticipate our realistic problem presentation (“The Seeker’s Dilemma”) will help boost IC security by diversifying HT detection methods. The key here is that it is unlikely that a single holy grail for HT detection will find HTs with a whole range of insertion criteria. Instead, the research community needs to generate diverse approaches to broaden HT detection. Each approach and detection idea (detection strategy) covers a portion of the problem space. 

The contributions of this work are as follows:
\begin{enumerate}
    \item We introduce The Seeker’s Dilemma, a formal definition of HT detection similar to finding HTs in real-world scenarios.
    \item  We propose a comparison flow for HT detection that is an update from legacy HT detection.
    \item We introduce \textit{Seeker1} which is a benchmark with potentially hidden HTs, to help improve HT detection.
    \item We use Principal Component Analysis (PCA) to analyze the significance of \textit{Seeker1} from a Machine Learning (ML) training perspective.
\end{enumerate}

The paper is organized as follows. Section~\ref{sec_threat_model} explains the threat model in our flow. Section~\ref{B} describes the existing state-of-the-art HT benchmarks and detection strategies. Section~\ref{H} introduces the notion of Hide\&Seek on a graph and how that problem is related to HT. Section~\ref{S_D} modifies the above problem to the new problem of “The Seeker’s Dilemma”. Next, Section~\ref{HT_B} describes the benchmark we have created, a version of “The Seeker’s Dilemma”. Section~\ref{R} provides results and analysis of our new benchmark with respect to the performance of three state-of-the-art HT detectors. Finally, Section~\ref{sec:discussion} discusses some caveats and constraints of our approach, and Section~\ref{C} concludes the paper.

\section{\change{Threat model}}
\label{sec_threat_model}
\change{The HT \textit{Hide\&Seek} game is played between two parties: 1) a perpetrator who embeds HTs and 2) a defender whose goal is to find them. The globalization of the silicon supply chain has provided countless opportunities for nefarious actors to compromise IC integrity. Attackers insert HTs to achieve at least one of these purposes: Denial-of-Service (DoS), leak secret information, and degrade performance~\cite{shakya2017benchmarking,gohil2022deterrent}. Several studies have comprehensively investigated the various silicon supply chain stages and the associated threats~\cite{shakya2017benchmarking,xue2020ten,li2016survey}. Although inserting HTs is possible in each of the stages, the difficulty varies from one stage to another. While malicious designers have the highest freedom to insert HTs in the design stage, this flexibility reduces at the foundry where the design has to be reversed-engineered first, and then HTs would be placed and routed on the layout~\cite{xue2020ten}. In this work, we assume that the attacker relies on Third-Party Electronic Design Automation (3P-EDA) tools to insert HTs at the synthesis stage\cite{xue2020ten}. The EDA tool enables the attacker to restructure the circuit’s netlist while maintaining the circuit functionality unchanged. It is worth mentioning that functional restructuring used in this study is inherently different from hardware obfuscation in the logic locking context~\cite{hoque2020hardware}. }


\change{ As for detection, we will use three detectors that assume the following threat models:
\begin{enumerate}
    \item HW2VEC assumes a rogue player in one of these scenarios: 1) the design team, 2) a 3PIP provider, or 3) 3P-EDA 
    \item DETERRENT assumes a perpetrator in a foundry capable of reverse engineering designs and embedding HTs
    \item Sarihi~\etal~\cite{sarihi2023trojanframework} assume the same threat model as DETERRENT 
\end{enumerate}}
\change{The HW2VEC threat model aligns well with our assumptions, whereas DETERRENT and the framework proposed by Sarihi \etal~\cite{sarihi2023multi} adopt a post-silicon logic testing threat model. Although test-based detectors are oblivious to functional restructuring when reporting detection accuracy, we use them to assess the stealthiness of the inserted HTs in \textit{Seeker1}. Our main focus and analysis, though, revolves around pre-silicon HT detection of netlists, primarily utilizing tools such as HW2VEC.}
\section{Background \& Related Work}
\label{B}

\change{In this section, we review the state-of-the-art HT detection methods (\ref{subsec:detection_background}), existing HT benchmarks and their shortcoming (\ref{subsec:HT_Bench_background}), followed by HT detection strategies in Game Theory (\ref{subsec:GameTheory}).} 

\subsection{HT state-of-the-art Detection}
\label{subsec:detection_background}
HT detection methods can be categorized based on their dependence on a golden model. If security engineers have access to a golden model or its specification, they will look for any deviations in the functionality of the devices from the expected standard behavior. 

Chakraborty~\etal~\cite{chakraborty2009mero} introduced MERO, a test vector generator for detecting HTs by stimulating rare-active nets multiple times. While it performs well against small circuits with rare triggers, its detection rate diminishes significantly as the circuit size increases. Lyu and Mishra introduced TARMAC~\cite{lyu2020scalable}, which maps the trigger activation problem to the clique cover problem and uses a SAT-solver to generate test vectors for each maximal satisfiable clique. Though it performs well, it lacks scalability and stability, and the implementation is not readily available for researchers~\cite{pan2021automated,gohil2022attrition}. 

TGRL~\cite{pan2021automated} is a reinforcement learning (RL) framework for hardware Trojan detection that uses a reward function to encourage activating as many rare nets as possible. Despite showing higher detection rates than MERO and TARMAC, it was not tested on any HT benchmarks. Sarihi~\etal~\cite{sarihi2023multi} proposed an RL-based multi-criteria HT detector that employs a tunable rewarding function. The authors generate test vectors utilizing three different rewarding functions and combine these detectors to enhance the ultimate detection accuracy~\cite{sarihi2023trojanframework}. DETERRENT~\cite{gohil2022deterrent} is another RL-based detection method that efficiently finds the smallest set of test vectors to activate multiple combinations of trigger nets. However, it focuses solely on signal-switching activities. It uses a SAT-solver to check compatibility, and its target detection domain is limited.

Machine learning (ML) approaches have also been used extensively towards HT detection~\cite{salmani2016cotdpaper,hasegawa2017trojan,yu2021hw2vec,gubbi2023hardware} when golden models are not available. Hasegawa~\etal~\cite{hasegawa2017trojan} proposed a random forest classifier trained on 51 circuit features extracted from Trusthub benchmarks to detect HTs. However, the method suffers from bias due to limited training data. HW2VEC~\cite{yu2021hw2vec} is a tool that converts hardware designs into other structures (\textit{i.e.}, dataflow graphs and abstract syntax trees) to extract structural features. The extracted features train a GNN (Graph Neural Network) for HT detection with a set of HT-free and HT-infected Trusthub circuits. COTD~\cite{salmani2016cotdpaper} uses unsupervised clustering analysis to detect HTs. The tool extracts “Sandia Controllability/Observability Analysis Program” SCOAP~\cite{goldstein1980scoap} parameters for each net and clusters circuits into HT-free and HT-infected. Notably, the mentioned tools depend on a single HT benchmark known as Trusthub, which limits the confidence about the performance of the methods on other datasets~\cite{gohil2022attrition,cruz2018automated}.

\subsection{HT benchmarks}
\label{subsec:HT_Bench_background}

Since HTs became a research interest, researchers have proposed various ideas to insert HTs in circuits. The benefit of having an inclusive HT benchmark is closely associated with developing robust HT detection methods and tools to thwart their impact. The first version of such attempts was available at Trusthub~\cite{salmani2013design,shakya2017benchmarking}, where several HT designs are available to study. Despite their valuable contribution to the HT research community, the benchmark lacks the size and variety needed to push the detection field forward~\cite{cruz2018automated,sarihi2022hardware,gohil2022attrition}. In particular, the problem is exacerbated when developing ML-based HT detectors. Such detectors need significant data (various HT-infected and HT-free circuits) to provide fair and unbiased results. An important problem observed in this space is the imbalance between HT-infected circuits versus HT-free ones, which negatively impacts the quality of the training data~\cite{liakos2022gainesis}. Subsequently, trained detectors can be biased towards favoring circuits as HT-infected, as we will see later in Section~\ref{subsec:HT_detection_analysis}. Trusthub benchmark contains log files that clearly explain the location and functionality of inserted HTs. 

To tackle Trusthub HT benchmarks shortcomings, various research endeavors have been made. Cruz ~\etal~\cite{cruz2018automated} presented a tool that inserts HTs in design netlists with configurable parameters such as the number of HT trigger nets, rare-net thresholds, number of HT instances, activation methods, types, and payloads. This work uses signal probability for HT triggers. The benchmark set generated by this tool is available on Trust-Hub~\cite{trusthub}. Yu~\etal~\cite{yu2019improved} proposed a similar tool that uses transition probability to identify rare nets, defined as the time required for the value of each net to toggle, and it is modeled using geometric distribution. This work suffers from the same shortcomings as~\cite{cruz2018automated}, where there is a single HT insertion strategy. We are curious if the generated benchmarks are available for download.

Gohil~\etal~\cite{gohil2022attrition} proposed an RL-based platform for HT insertion called ATTRITION. The authors assume reverse-engineering capabilities for the attacker to construct HT triggers where the target is signal probability. The agent’s objective is to identify a set of “compatible” rare nets, \textit{i.e.}, a group of rare nets that can be activated together by the same input test vector generated by a SAT-solver. The generated benchmarks have a structural format very similar to Trusthub’s.

Liakos~\etal~\cite{liakos2022gainesis} employed GANs (Generative Adversarial Networks) to replicate the characteristics of HT-free and HT-infected circuits characteristics, such as structural, power, and timing features. The study utilizes a \textit{feature generative approach} with GANs to generate synthetic data for training more effective HT detectors. Despite its novelty, the tool heavily depends on Trusthub and only reproduces data patterns similar to Trusthub, lacking the introduction of new functional dimensions.

Sarihi~\etal~\cite{sarihi2022hardware} uses RL to develop a tunable HT insertion tool that inserts HTs in design netlists. Netlists are first converted to graphs, and an RL agent interacts with the circuit environment through various actions that allow the agent to move the trigger inputs throughout the ISCAS-85 circuits, exploring various potential locations suitable for HT embedding. Trigger selection is based on a combination of SCOAP parameters. The benchmark follows a structural Verilog format. Despite the insertion flexibility, the benchmark was not tested against any HT detection techniques.

\subsection{\change{Game Theory for HT detection}}
\label{subsec:GameTheory}


\change{Kamhoua \etal~\cite{kamhoua2016game} is the first to study the HT dilemma as a zero-sum game between an attacker and a defender. The goal is to use the Nash Equilibrium (NE) to find the optimal test batch that reveals HTs in the presence of an intelligent
attacker. The study claims robustness against irrational attacker scenarios as well.  Saad \etal~\cite{saad2017hardware} consider uncertainty and risk in decision-making in both the attacker’s and defender’s behaviors by applying principles from \textit{prospect theory} (PT)~\cite{kahneman2013prospect}. Das~\etal~\cite{das2020think} also uses PT but with a slight tweak in the game. The defender learns the attacker’s strategies in the ``learning stage'' and the defender uses this knowledge in the ``actual game''. This enables the attacker to ``play dumb'' and deceive the defender in the learning phase.  Brahma \etal~\cite{brahma2021game} study HT testing by setting a budget constraint on the testing process while dealing with single and multiple HT types. Later, Nan \etal~\cite{nan2023game} improved this work by adding human errors and biases in the insertion and detection processes. Gohil \etal~\cite{gohil2021games} investigate the attack and defense strategies in a split-manufacturing environment. The authors analyze two different games with five attack and nine defense strategies.}

\change{Despite the novelty in the literature, none of the proposals study HT insertion and detection through a Hide\&Seek lens. In the next Section, we will elaborate on the details of such a game as it relates to HT detection.}

\section{Hide\&Seek in Cybersecurity}
\label{H}

Earlier work has investigated and helped define the ideas of \textit{Hide\&Seek} from a mathematical perspective. A broad view of these mathematical games exists by Steven Alpern and Shmuel Gal’s work~\cite{alpern2006theory} on Search Games, which, itself, was built on the early groundwork of Isaac’s Differential Games~\cite{isaacs1965differential}. Similarly, the idea of search and optimal search has been deeply explored and surveyed by Stone~\cite{stone1976theory}. 

Inspired by the Hide\&Seek game defined by Chapman \etal~\cite{chapman2014playing}, we define \textit{Hide\&Seek} as a zero-sum game on a graph or network. The goal of the \textit{Seeker} $S$ is to find at least one node where the \textit{Hider} $H$ has hidden objects. We define this game $(H, S)$, and it is said to have length $L$ if \textit{Seeker} makes exactly $L$ guesses before finding all nodes where the \textit{Hider} has hidden objects. In adversarial \textit{Hide\&Seek}, which is the typical way most children and adults would play the game, $H$ tries to maximize $L$ and, conversely, $S$ tries to minimize $L$. The graph the game is played upon is defined as $G = (V, E)$, where $V$ is a set of vertices (nodes), and $E$ is a set of edges, such that $E \subseteq V \times V$. Let the number of vertices be defined as $n = |V|$. Chapman \etal{} define the edges of this graph to have weights such that there is an explicit cost associated with its traversal, \textit{i.e.}, $Cost: E \rightarrow {0, ..., c}$ where $c$ is an upper bound for the weights. 
The idea of movement in seeking can be replaced by the idea that the \textit{Seeker} can teleport anywhere in the graph and query if something is hidden at a cost of $1$ per guess.

In the game, the \textit{Hider} can conceal a series of objects, which is the set $\mathcal{H}$ (where $\mathcal{H} \subseteq V$). Let $k = |\mathcal{H}|$ be the number of hidden objects selected according to the \textit{Hider’s} strategy. In the game, the \textit{Hider} hides $k \geq 1$ objects, and the \textit{Seeker} can search for those objects in $L$ queries or moves. From our perspective, a query can be a selection of any vertices $V$ by any means possible, and the cost details depend on the practical search mechanism. 

Finally, we direct the reader to the importance of Chapman \etal{} work~\cite{chapman2014playing} and highlight that their work was the first, to our knowledge, to relate cybersecurity and \textit{Hide\&Seek} as a fundamental connection between cybersecurity and theory, and in this vein, we extend this connection to the hardware security domain.

\section{HT Detection: A Cornered Case Hide\&Seek}
\label{S_D}

\subsection{Digital Circuits Representation}

To better understand the relationship between digital circuits and where HTs can be hidden, we start by providing a simple definition of a digital circuit as a directed graph $G = (V, E)$, where:
\begin{itemize}
\item $V$ is a set of vertices representing logic gates.
\item $E$ is a set of edges representing connections between gates.
\item Each vertex in $V$ has a corresponding Boolean function that defines its output based on the inputs to the gate.
\item A subset of vertices in $V$ are designated as primary inputs, denoted by $I$.
\item Another subset of vertices in $V$ are designated as primary outputs, denoted by $O$.
\end{itemize}
\change{This definition preserves essential information from the circuit netlist, aligning effectively with the assumption of functional HTs in our threat model.}


An HT can be represented as a vertex $t \in V$, inserted into the original digital circuit by an attacker with malicious intent. This vertex $t$ is connected to the existing circuitry by a set of edges $E_{HT}$. The attacker’s adoption of $t$ and $E_{HT}$ represent the malicious intention of the inserted HT. Typically, the HT is inserted to replace existing edges in the original graph $G$. For an inserted HT, the set of primary inputs $I$ and primary outputs $O$ remain the same as for the original circuit; however, the presence of the HT means that the functional logic behavior of the circuit may no longer be trusted. The malicious behavior of the HT can be triggered by a specific input or internal state, causing it to activate and perform the intended malicious action. For example, a triggered HT might cause an incorrect calculation internally in the circuit as a denial of service attack.

\begin{figure}[t]
  \centering
  \includegraphics[scale=.48]{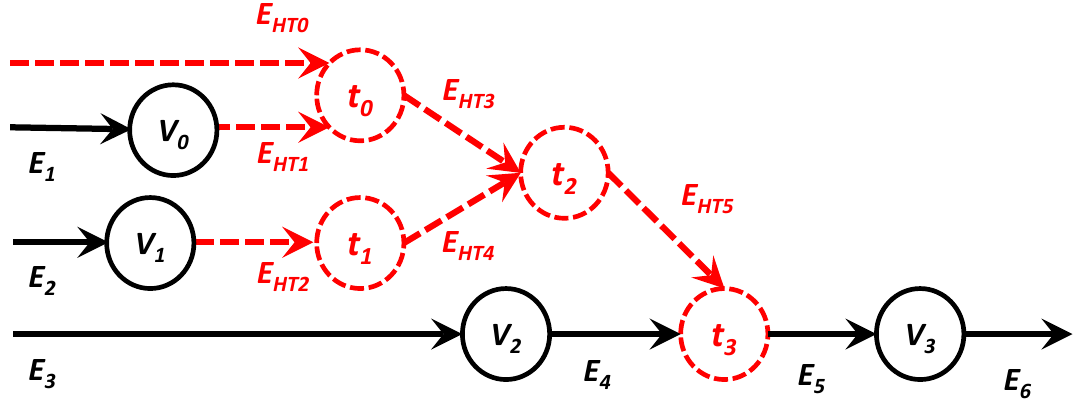}
  \caption{$G_{HT}$, shown in red, is embedded in $G_{T}$.}
  \vspace{-3mm}
  \label{fig:HT_graph}
\end{figure}

The HT $t$ can also be larger than a single vertex and can be defined as a circuit itself such that $t = (V_t, E_t)$ and the primary inputs and outputs of $t$ are connected to the circuit under attack via edges $E_{HT}$. \change
{Finally, we can define the graph $G_{HT}$ representing a digital circuit with an inserted HT as 
$G_{HT} = (V \cup V_{t}, E \cup E_{HT} \cup E_{t})$.}~Figure~\ref{fig:HT_graph} shows an HT (colored red) inserted into a small circuit with the above graph definitions. 

\subsection{The Seeker’s Dilemma}

The problem of traditional \textit{Hide\&Seek} on a graph once applied to the HT detection problem (a situation which we call ``The Seeker’s Dilemma'') is that given an IC or netlist from the fabrication process, the \textit{Seeker} ($S$), does not know whether or not an HT has been hidden.

Therefore, the \textit{Seeker’s Dilemma} is a redefinition of \textit{Hide\&Seek} with one fundamental condition, \textit{i.e.}, a two-player game has $H$ the \textit{Hider} and $S$ the \textit{Seeker}. As defined earlier, $k = |\mathcal{H}|$ is the number of objects hidden, and in the Seeker’s Dilemma, $k$ hidden objects have the condition \underline{$k \geq 0$}. Additionally, \underline{the value of $k$ is unknown by the \textit{Seeker}}. Adding this condition and the unknown information transforms the problem into what we define as ``The Seeker’s Dilemma'', and from a complexity theory perspective, makes the problem significantly harder from a real-world perspective. 

The Seeker’s Dilemma has very different strategical implications such that the \textit{Hider} hides $k$ objects (could be nothing if $k=0$) and the \textit{Seeker} ($S$) searches for hidden objects in $L$ queries or moves, where $H$ with a $k \geq 1$ tries to maximize $L$ and, conversely, $S$, tries to minimize $L$. When $k = 0$, $S$ spends minimal time seeking what is yet to be hidden (returns after one trial, $L = 1$).   

\subsection{Benchmarking}
\label{subsec:benchmarking}

From a security engineer perspective, the question is, given that $k \geq 0$, how can I efficiently spend my time finding HTs when they exist and minimize my effort when they do not?

As previously proposed in~\cite{sarihi2023multi}, this brings us to the four conditions in which an HT detector can classify a circuit.

Figure~\ref{fig:TFPN} shows four possible outcomes when an HT detection tool studies a given circuit. From the tool user’s perspective, the outcomes are probabilistic events. For example, when an HT-free circuit is being tested (there is no HT and $k = 0$), the detection tool may either classify it as an infected or a clean circuit, \textit{i.e.}, $Prob(FP) + Prob(TN) = 1$ where $FP$ and $TN$ stand for \textit{False Positive} and \textit{True Negative} events. Similarly, for HT-infected circuits ($k \geq 1)$, we have $Prob(FN) + Prob(TP) =1$. 

\begin{equation}
    \label{equ:metric}
    Conf. \, Val = \frac{(1-FP)}{(1/\alpha+FN)}
\end{equation}

$FN$ and $FP$ are two undesirable outcomes where detectors misclassify a given circuit. The $FN$ detection cases pose a greater danger as they result in a scenario where we verify an HT-infected chip. So, we need to know how HT detection tool users prioritize $FN$ and $FP$ cases. We define a parameter $\alpha$ as the ratio of the undesirability of $FN$ over $FP$. The tool user determines $\alpha$ based on characteristics and details of the application that eventually chips will be employed in, \textit{e.g.}, the risks of using an infected chip in a device with a sensitive military application versus using a chip in a home appliance. Note that the user sets this value, which is not derived from the actual $FP$ and $FN$. After $\alpha$ is set, it is plugged into Equation~\ref{equ:metric} and a general confidence basis $ Conf. \, Val $ is computed. 

\subsection{\change{HT Detection Complexity}}
\label{detection-complexity}
\change{Inserting and detecting HTs are similar in an aspect, \textit{i.e.}, selecting a set of rare nets for insertion purposes or narrowing down the detection search space. While even semi-optimal trigger selection could still lead to hard-to-detect HT instances, the defender, on the other hand, faces major challenges. Here, we list a few:}

\change{1) Exhausting all the possible test vectors and comparing them against the golden model is not feasible. The defender mostly relies on heuristic algorithms to narrow the search space and speed up the detection process. }

\change{2) Kamhoua~\etal~\cite{kamhoua2016game} states that inserting no Trojans can be an effective strategy by an attacker, making the defender wonder if an HT has been inserted in the circuit. This uncertainty is represented as a probabilistic distribution that might result in false positive classifications. Incorrect classification leads to distrust in the chip source and engenders financial damage and reputation loss for the vendor.}

\change{3) Defender’s job gets significantly harder when detecting HTs with various trigger widths. This characteristic exponentially increases the attacker’s insertion opportunities (different trigger combinations). Even if the defender can guess the rare net set from the attacker’s perspective, it has to deal with multiple insertion strategies (multiple rare net sets) to ensure enough bases have been covered. Moreover, a rational attacker would not always follow the insertion strategy with the highest payoff, which further convolutes the defense strategy~\cite{kamhoua2016game}. It is worth mentioning that attackers might not exceed the trigger width threshold due to side-channel implications. This will depend on parameters such as the circuit size and power consumption.}

\change{From the \textit{Seeker’s} perspective, HTs are hidden in a directed acyclic graph with \textit{V} vertices and \textit{E} edges. The \textit{Hider} can use any graph edges to build the HT triggers. The total number of HTs with $q$ trigger width (number of HT inputs) can potentially be $\binom{E}{q}$; we assume that the attacker can insert HTs by adopting any of the $n$ different strategies. Also, the attacker decides how many of the HT’s trigger inputs are from the set of rare nets (edges), \textit{i.e.}, a sub-set of trigger inputs can be chosen from a set of rare nets $(r)$ and the rest from regular nets $(g)$. Let us assume that the attacker picks $r_i$ and $g_i$ for rare and regular nets under the $ith$ strategy. Accordingly, the \textit{Seeker} faces the complexity in Equation \ref{complexity}}. \change{The \textit{Seeker} must choose a set of rare nets}

\begin{equation}
\label{complexity}
    \color{black}\sum_{q=2}^{M}\sum_{p=0}^{q}\sum_{i=1}^{N} \binom{r_i}{p}\times\binom{g_i}{q-p}
\end{equation}
\change{where $q$ enumerates the number of trigger inputs (with the maximum of $M$), $p$ enumerates the number of trigger inputs selected from the set of rare nets (with the maximum of $q$), and $i$ enumerates the strategy used (with the maximum of $N$). As inferred from the equation, this problem has factorial time complexity.}

\section{Benchmark for HT Detection}
\label{HT_B}

The major challenge with existing HT benchmarks is that these circuits are known-knowns in terms of the existence of HTs, \textit{i.e.}, almost in all cases, $k = 1$. This means that both the inserting and detecting sides know the situation, which we believe oversimplifies the problem. Despite being very helpful to the community, Trusthub benchmarks~\cite{trusthub} fall into this class. To address this gap, we have created The Seeker’s Dilemma HT benchmark (we call it \textit{Seeker1}) featuring $k \geq 0$. The \textit{Seeker1} benchmark suite is available at~\cite{sarihi2023multi}.

\begin{figure}[!t]
    \centering
\includegraphics[scale=0.60]{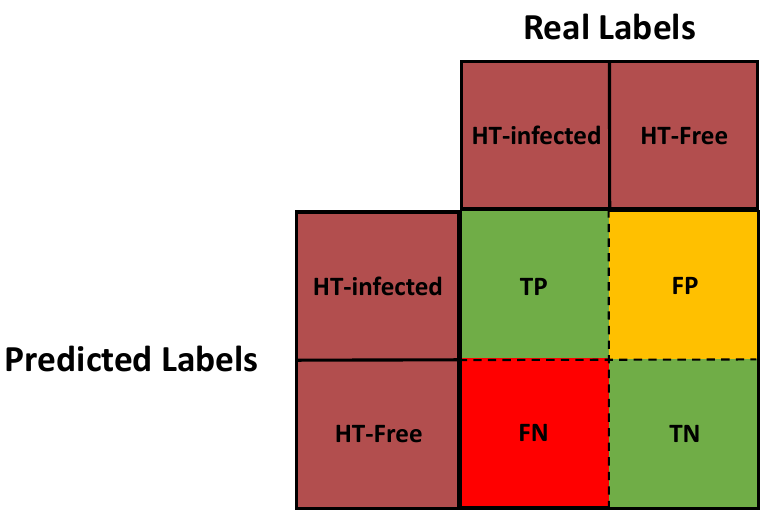}
    \caption{Possible outcomes of an HT detection trial.}
    \vspace{-4mm}
    \label{fig:TFPN}
\end{figure}

The goal is to create a benchmark that allows for each of the four cases ($TP$, $TN$, $FP$, $FN$) as defined above. We have initially used $8$ clean circuits. For the $i^{th}$ clean circuit ($1 \leq i \leq 8$), referred to as $Golden_{i}$, we map that circuit with several \change{CAD transformation techniques that keep the circuits’ functionally unchanged} (described later in this section). We have used $nb$ different \change{functional restructuring} techniques to create a set of netlists $B_{i} = \{b_{i_{1}}, b_{i_{2}} \ldots, b_{i_{nb}}\}$ (where $nb = |B_{i}|$) featuring that each of the \change{functionally transformed} versions may or may not contain an HT. This means that for each \change{functionally transformed equivalent circuit}, $k$ is either $0$ or $1$. For HT detection, any HT detection tool should evaluate and present its identification with which $B_{i_{j}}$ circuits are HT infected. This identification list (in a specified format) is submitted to a judge (Figure~\ref{fig:benchmark}), who then sends back results that will give statistics on how the tool performed on the benchmark in all four categories of detection shown in Figure~\ref{fig:TFPN}.

Benchmarking in any domain has several challenges for a community~\cite{jamieson2018benchmarking, jamieson2010benchmarking}, 
and HT detection needs similar considerations. For our proposed approach above, we provide a Python script 
that for a given benchmark prepares a report for a detection test submitted to the judge. This, however, has potential problems. For example, how should the benchmarks be formatted? Who is a fair judge? How many submissions can an HT detector perform, and how quickly can resubmissions happen?

In cybersecurity, we propose the following to address some of these problems. First, we propose that all benchmarks $B_{i_j}$ be released in a netlist-based Verilog structural form consisting of gate-level models using the combinational logic set {BUF, \textit{NOT}, \textit{AND}, \textit{OR}, \textit{XOR}, \textit{NAND}, \textit{NOR}, XNOR}.  However, the golden model ($Golden_{i}$) would be best provided as an RTL Verilog design, allowing researchers to see the higher-level structure and purpose of the benchmark, but a clean netlist model is acceptable. This allows for a consistent format for combinational circuits but does not deal with sequential circuits\footnote{Currently, our benchmarks focus on combinational circuits, and we have not come up with a consistent solution for sequential circuits. This will be addressed in our future work.}. Second, we suggest a distributed set of benchmarks for judging where research groups worldwide create their own benchmarks for the Seeker’s Dilemma where they know whether their $B_{i_j}$ is infected with an HT. Each judge would then be required to run the reports for each of their released benchmarks every month for all submitted detection runs (automated by our Python script). We propose that a benchmark set would have a three-year supported life span where the internal knowledge (of what is infected and what is clean) would be released openly after the three years. In this way, time is used to allow for fair benchmarking in that detectors cannot be tuned to the benchmarks by repeated attempts (submitting results in a brute-force manner is not allowed to crack the benchmarks unfairly), nor can a single judge provide favorable results.

\subsection{Benchmark Creation}
\label{subsec:obf}

Without loss of generality, we have selected the original circuits for \textit{Seeker1} from ISCAS-85’s~\cite {bryan1985iscas} combinational designs that have been used throughout CAD research community~\cite{cruz2018automated,reshma2019hardware,gohil2022attrition,gohil2022deterrent,yu2021hw2vec,jyothi2017taint,chakraborty2009mero,pan2002teaching,lyu2020scalable}. This benchmark has been widely used to evaluate the effectiveness of logic synthesis tools, technology mapping algorithms, test generation algorithms, timing analysis, various optimization techniques for digital circuits, and HT detection. So, we believe that ISCAS-85 is the first candidate to generate \textit{Seeker1}. 

\begin{figure*}
    \centering
    \begin{subfigure}{0.48\textwidth}
        \centering
        \includegraphics[width=\linewidth]{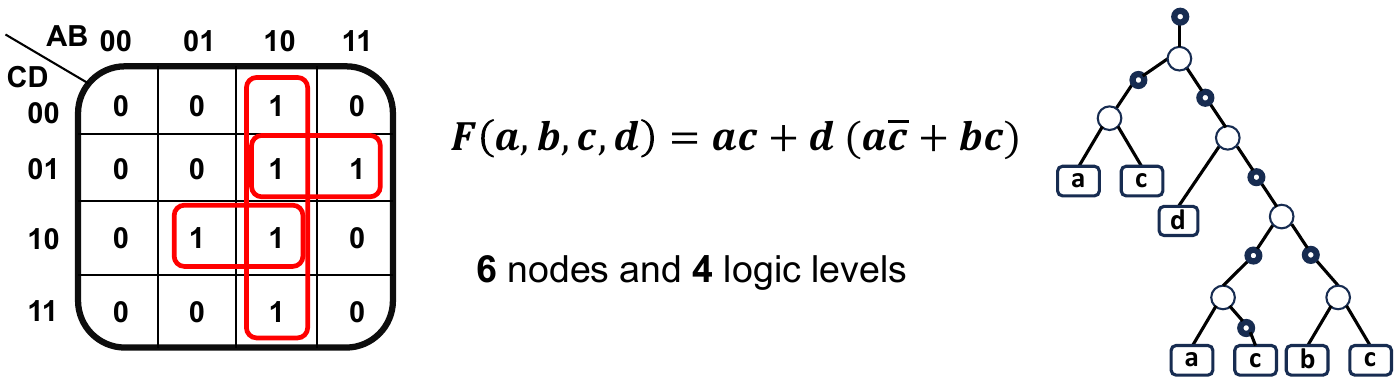}
        \caption{AIG representation \#1}
        \label{fig:subfig_AIG_1}
    \end{subfigure}
    \hfill
    \begin{subfigure}{0.48\textwidth}
        \centering
        \includegraphics[width=\linewidth]{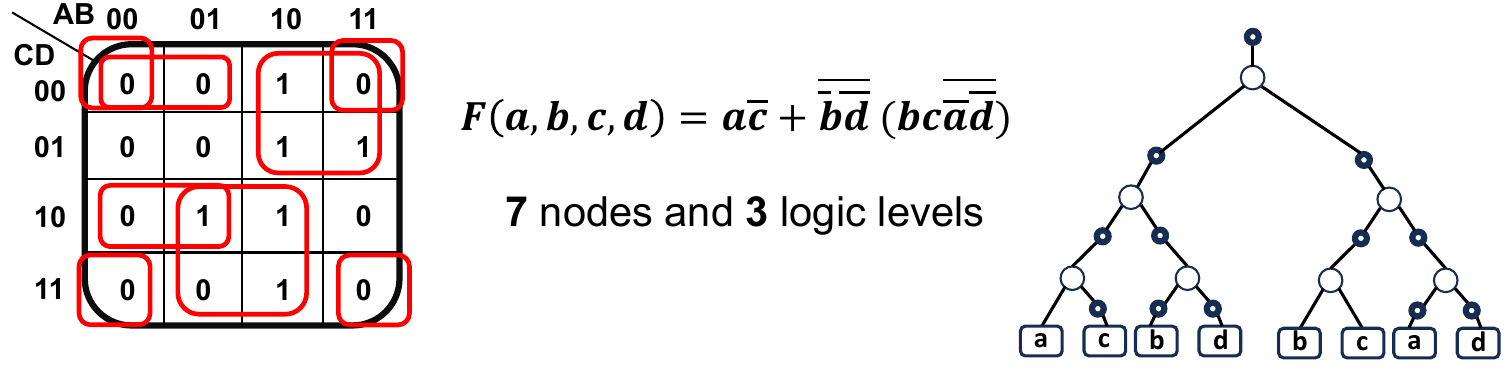}
        \caption{AIG representation \#2}
        \label{fig:subfig_AIG_2}
    \end{subfigure}
    \caption{Two representations of a circuit with the same truth table. Representation \#1 is aimed at improving the area in terms of comparatively fewer nodes, while representation \#2 enhances delay with fewer logic levels}
    \label{fig:AIG}
    \vspace{-6mm}
\end{figure*}

Boolean functionality of digital circuits can be described in various formats, \textit{e.g.}, Sum-of-Products (\textit{SOP}), and Products-of-Sum (\textit{POS}). SOP and POS have a \textit{canonical sum} that is a unique representation. The canonical form is a unique representation of each circuit; it enables ease of implementation with \textit{AND}, \textit{OR}, and \textit{NOT} gates for every digital circuit. Another logic format is the AND-Inverter Graph (AIG), where circuits are represented with two-input \textit{AND} gates and \textit{NOT} gates~\cite{mishchenko2018integrating}. This format can be derived from a circuit by factoring the functions of the logic nodes and converting the AND/OR gates to two-input ANDs and NOTs using DeMorgan’s rule~\cite{mishchenko2018integrating}. Various algorithms have employed AIG circuit representations to enhance the area, optimize delay in synthesis, and expedite formal equivalence checking procedures~\cite{chowdhury2021openabc}. AIGs are not canonical, \textit{i.e.}, a boolean function has various AIG representations~\cite{mishchenko2005fraigs}. Figure~\ref{fig:AIG} shows two of these formats where the AIG in Figure~\ref{fig:subfig_AIG_1} has less area and more delay than Figure~\ref{fig:subfig_AIG_2}.   

We use ABC~\cite{brayton2010abc} (an open-source logic optimization tool) and employ its various representations and logic optimization techniques to alter the structure of the original ISCAS-85 circuits, thereby hiding the existence of HTs and rendering their detection through visual inspection or simple comparison techniques hard. We employ $18$ \change{functional restructuring} methods that utilize one or more of the following techniques:

\begin{itemize}
\item \textbf{Structural Hashing:}
A one-level structural hashing technique converts the circuit to its AIG equivalent. It ensures the AIG is stored in a compact form~\cite{mishcenko}

\item \textbf{Balancing:} This algorithm takes a circuit in the AIG format and aims at generating an equivalent AIG with the minimum possible delay. Delay is the time taken for a signal to traverse through the logic circuit, and the technique minimizes delay by reducing the logic level of the circuit.

\item \textbf{Rewriting:} The algorithm tries to reduce the number of AIG nodes and logic levels by replacing sub-circuits with smaller recomputed sub-graphs.
    
\item \textbf{Resubstitution:} The algorithm identifies sub-circuits that can be simplified and replaced by a simpler Boolean representation.

\item \textbf{Refactoring:} ABC carries out iterative logic cone collapsing and refactoring in the AIG representation, aiming to minimize the number of AIG nodes and logic levels.

\item \textbf{Gate-Size change:} Two input gates in AIG are transformed into multi-input \textit{AND} gates.

\item \textbf{Fraiging:} The tool converts the existing network into a functionally reduced AIG, a modified version of AIGs where each node possesses unique functionality compared to other nodes in the AIG.
\end{itemize}

We do not reveal the details of our \change{functional restructuring} steps to prevent reverse-engineering to regenerate our benchmarks and easily determine if an HT exists. ABC functionally transforms the circuit such that, unlike Trusthub benchmarks where the HT circuit is discernible in the HDL code,  the HT payload and trigger logic are hidden in the Verilog file.

Depending on the threat model, the Seeker’s Dilemma benchmark might need another approach to \change{functionally transform}, hide, and create the benchmark. The above approach is reasonable for creating a netlist-based benchmark. With this approach, we create a benchmark with 8 of the ISCAS-85 combinational benchmarks ($c880$, $c1355$, $c1908$, $c2670$, $c3540$, $c5315$, $c6288$, and $c7552$). For each of these $8$ golden circuits, we add HTs to some of them (again, we do not release this number, noting it is different for each circuit in the benchmark).

\subsection{The Baseline Benchmark}
\label{subsec:HT_insertion}

We use the HT circuits generated by an RL framework explained in~\cite{sarihi2022hardware,sarihi2023trojanframework}. The tool automatically inserts HTs with various criteria. We pick 100 inserted HTs from each of $c880$, $c1355$, $c1908$, $c2670$, $c3540$, $c5315$, $c6288$, and $c7552$. \change{The characteristics of each circuit are discussed in ~\cite{bryan1985iscas} in depth}. Next, we employ the \change{functional restructuring} techniques explained in the previous section to \change{functionally transform} and hide the HT-infected circuits. We apply the same functional restructuring techniques to ISCAS-85 HT-free circuits and create 18 versions of each. 

\section{Analysis of our Benchmark}
\label{R}

To analyze our first Seeker’s Dilemma benchmark, \textit{Seeker1}, we test the benchmark against existing HT detection tools.
We use three different HT detection strategies/tools to measure the quality of inserted HTs. We run the test vectors developed in~\cite{sarihi2023multi,sarihi2023trojanframework}, which we will call RL\_HT\_DETECT, the test vectors from DETERRENT proposed in~\cite{gohil2022deterrent}, and the open-source HW2VEC~\cite{yu2021hw2vec}. All three tools are ML-based HT Detectors and can be trained and tuned differently. Despite using ML in their backbones, the three tools differ in their detection strategies. 

RL\_HT\_DETECT and DETERRENT are test-based HT detectors, \textit{i.e.}, a golden model is required for HT detection (threat model discussed in Section~\ref{sec_threat_model}). Although finding a golden model to compare against is not always a trivial task, these detectors are resilient against \change{functional restructuring} techniques since \change{functional restructuring} does not change the functionality of circuits but instead changes their structure. Moreover, these detectors do not yield any $FPs$.
HW2VEC, on the other hand, is an open-sourced GNN-based HT detector that extracts behavioral features from hardware designs and generates a dataset with $200$ features. The dataset is then used to train a binary classifier, which reports all four classification cases explained in Section~\ref{subsec:benchmarking}. Despite its novelty in learning circuit structures, the detector’s performance depends on the quality and quantity of the dataset on which it is trained. Techniques such as leave-one-out cross-validation are used in most ML-based HT detectors where training data is scarce~\cite{hasegawa2017trojan}; however, it does not solve the data shortage problem entirely.
 
We use RL\_HT\_DETECT in the \textit{Combined} mode~\cite{sarihi2023trojanframework} where the three detection strategies are combined together. We use the open-sourced test vectors from DETERRENT and RL\_HT\_DETECT to report detection rates. We also train an MLP (multi-layer perceptron) network to classify the generated graph embeddings from HW2VEC with the same configuration described in~\cite{yu2021hw2vec}. The classifier is trained  with two training datasets described as the following:

\begin{enumerate}
\item The \textit{TJ\_RTL} dataset used in~\cite{yu2021hw2vec} for training HW2VEC. This dataset mostly contains communication protocols and encryption algorithms from \url{Trusthub.org}. The dataset contains $\textbf{26}$ HT-infected and $\textbf{11}$ HT-free instances. The rest of the paper refers to the HT detection reports under this scenario as \textit{$S_1$}.

\item We add two versions of ISCAS-85 HT-free benchmarks~\cite{cruz2018automated,hansen1999unveiling} to \textit{TJ\_RTL} to make the dataset labels more balanced. This action adds $\textbf{16}$ more HT-free instances to the previous set. We refer to the HT detection reports under this scenario as \textit{$S_2$} hereafter.
\end{enumerate}

We do not use any of the \change{functionally transformed} data during the training process of the MLP to mimic real-world situations under which neural networks with trained weights and biases are used to predict the labels of unseen data. In ML, this process is called \textit{inference}~\cite{chollet2021deep}.


\subsection{Principal Component Analysis}

To simplify the illustration of each dataset instance and convert it to a more human-readable fashion, we use Principal Component Analysis (PCA)~\cite{bro2014principal}. PCA is used to decrease data dimensionality from $200$ attributes to a handful of principal components that collectively explain a significant portion of the total variance. The PCA algorithm identifies principal components, \textit{i.e.}, orthogonal axes along which the data varies the most. PCA analysis is already used for HT detection based on side-channel information~\cite{liu2022pca,shende2016side,wang2013malicious}.

\begin{figure}[!t]
\centering
\includegraphics[width=\linewidth]{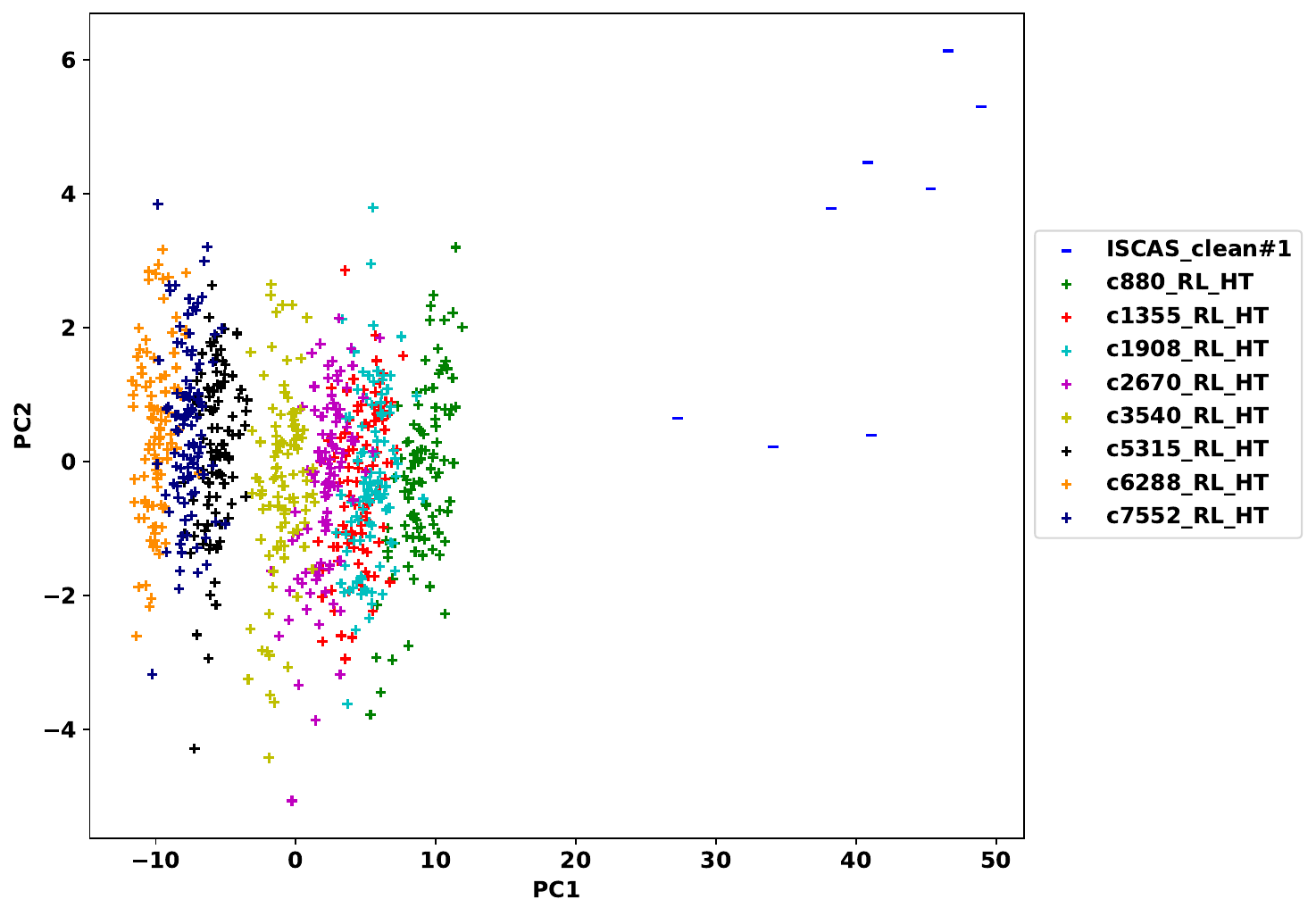}
\vspace{-5mm}
\caption{The distribution of clean ISCAS-85 circuits vs. RL HTs generated by~\cite{sarihi2022hardware,sarihi2023trojanframework} after applying PCA analysis}
\label{fig:RL_HT_vs_clean}
\vspace{-5mm}
\end{figure}

In Figure~\ref{fig:RL_HT_vs_clean}, the layout of ISCAS-85 circuits is visualized both before and after HT insertion (refer to Section~\ref{subsec:HT_insertion}), with PCA analysis applied. The $x$ axis corresponds to PC1 (Principal Component 1), and the $y$ axis represents PC2. The plots use negative markers (\textbf{-}) to represent HT-free circuits and positive markers (\textbf{+}) to depict HT-infected instances. This convention is followed in the rest of the paper. As shown in Figure~\ref{fig:RL_HT_vs_clean}, the HT-free instances have an entirely different pattern than the HT-infected ones. Surprisingly, a simple linear classifier can classify the data correctly in this case; however, the pattern is not as observable for hidden HT instances. 

\begin{figure}[!t]
\centering

\begin{subfigure}{0.48\textwidth}
\centering
\includegraphics[width=\linewidth]{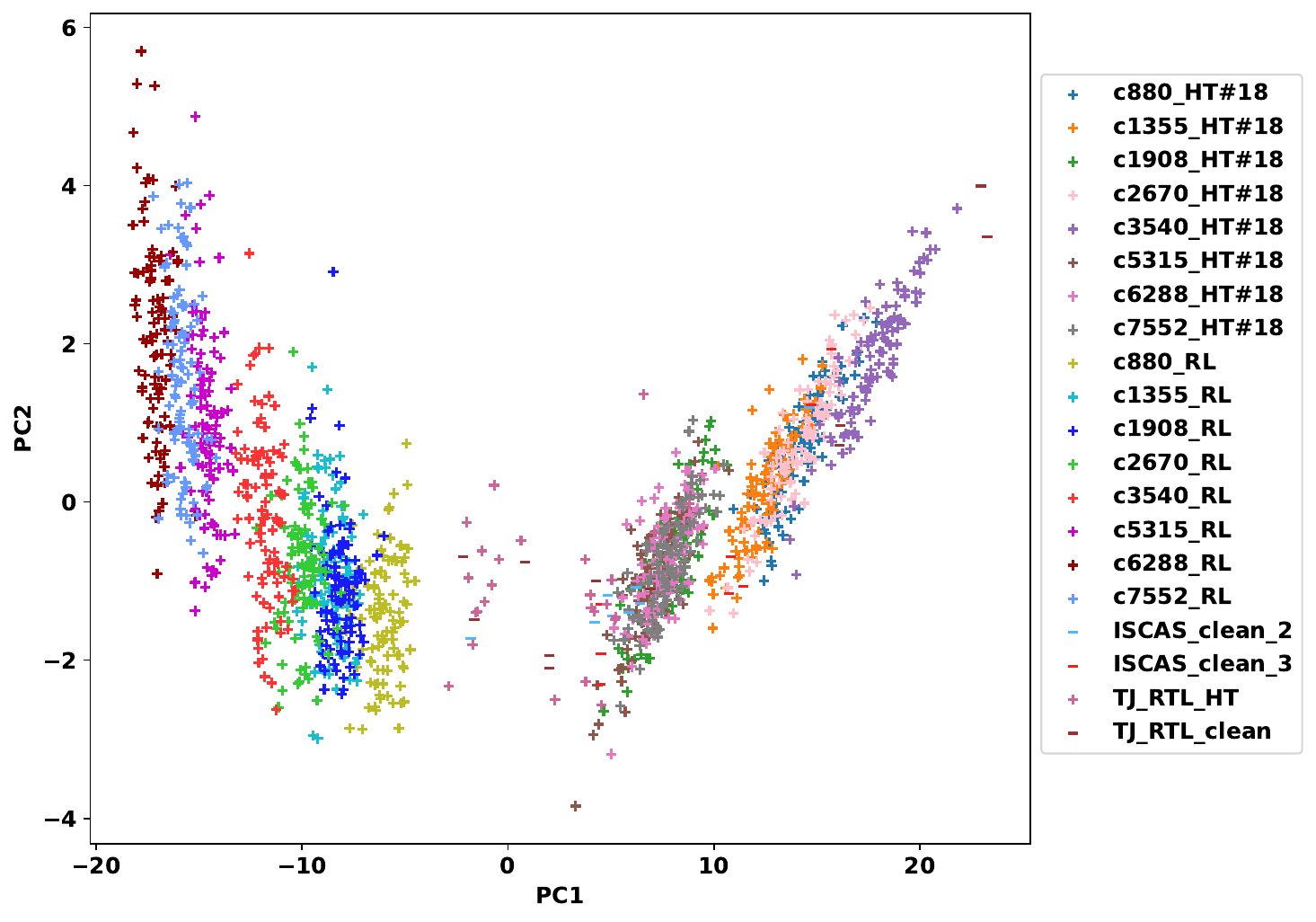}
\vspace{-4mm}
\caption{PC1 vs.\ PC2}
\label{subfig:RL_vs_ABC_1_2}
\end{subfigure}

\vspace{1mm}

\begin{subfigure}{0.48\textwidth}
\centering
\includegraphics[width=\linewidth]{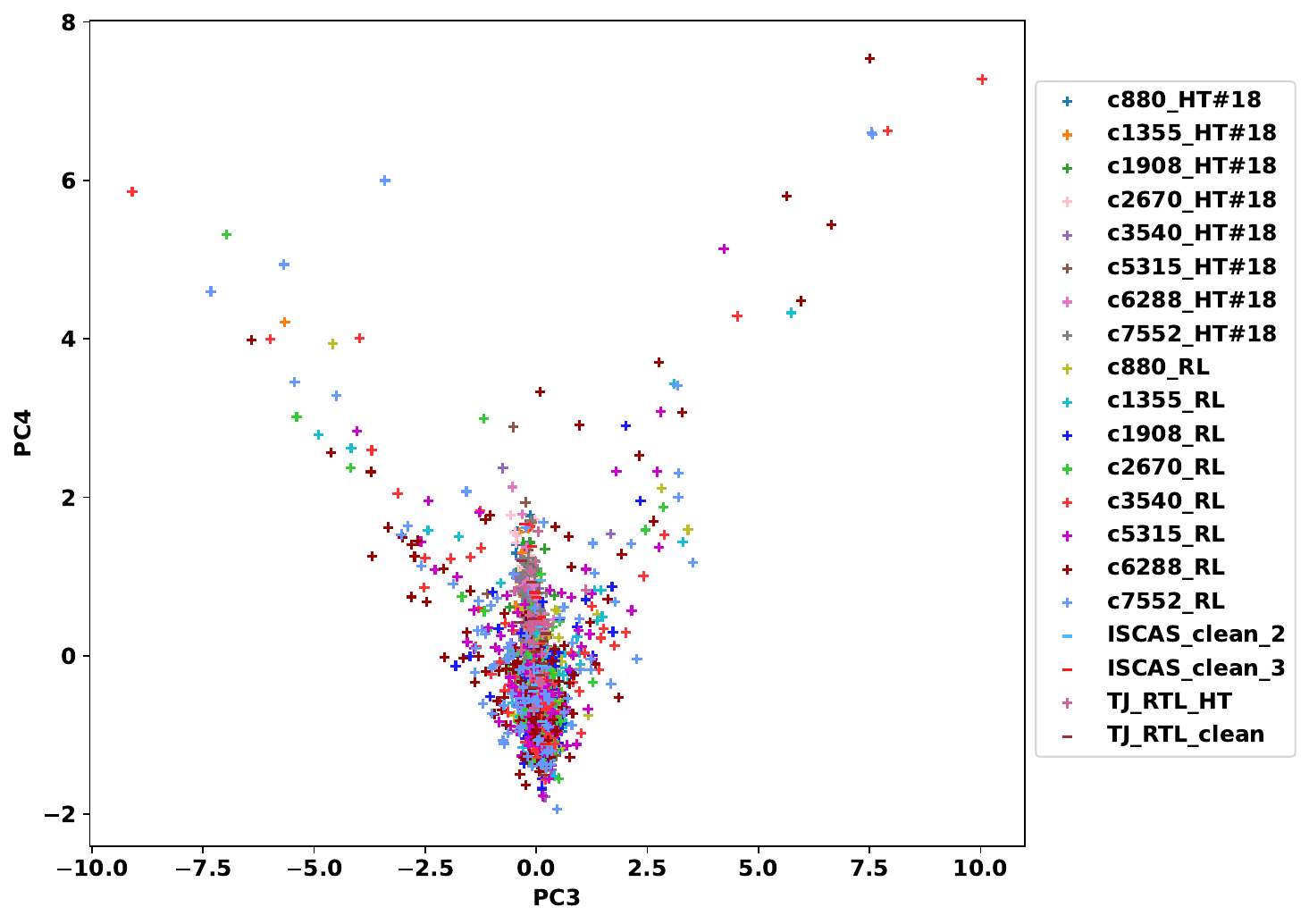}
\vspace{-4mm}
\caption{PC3 vs.\ PC4}
\label{subfig:RL_vs_ABC_3_4}
\end{subfigure}
\vspace{-2mm}
\caption{The layout of various HT-infected and HT-free circuits on the PCA plot
}
\label{fig:RL_vs_ABC}
\vspace{-6mm}
\end{figure}

Figure~\ref{fig:RL_vs_ABC} plots the distribution of RL-inserted HTs, the \textit{TJ\_RTL} dataset, two versions of ISCAS-85 HT-free benchmarks from~\cite{cruz2018automated} and~\cite{hansen1999unveiling}, and our hidden HTs using the \change{functional restructuring} method number-18. As can be seen from Figure~\ref{subfig:RL_vs_ABC_1_2}, the hidden HTs (right side of the figure) have a different layout compared to the original RL HTs (left side of the figure). The \textit{TJ\_RTL} data points are mostly observable around $\textit{PC1=0}$. We also plot PC3 vs.\ PC4 in Figure~\ref{subfig:RL_vs_ABC_3_4}; however, no immediate patterns or clusters can be recognized. The data layout raises an interesting question: ``How accurately can a detector trained on the Trusthub dataset classify HT-infected and HT-free circuits?'' We will try to investigate this further in Section~\ref{subsec:HT_detection_analysis}. 

\begin{figure}[!t]
\centering
\includegraphics[width=\linewidth]{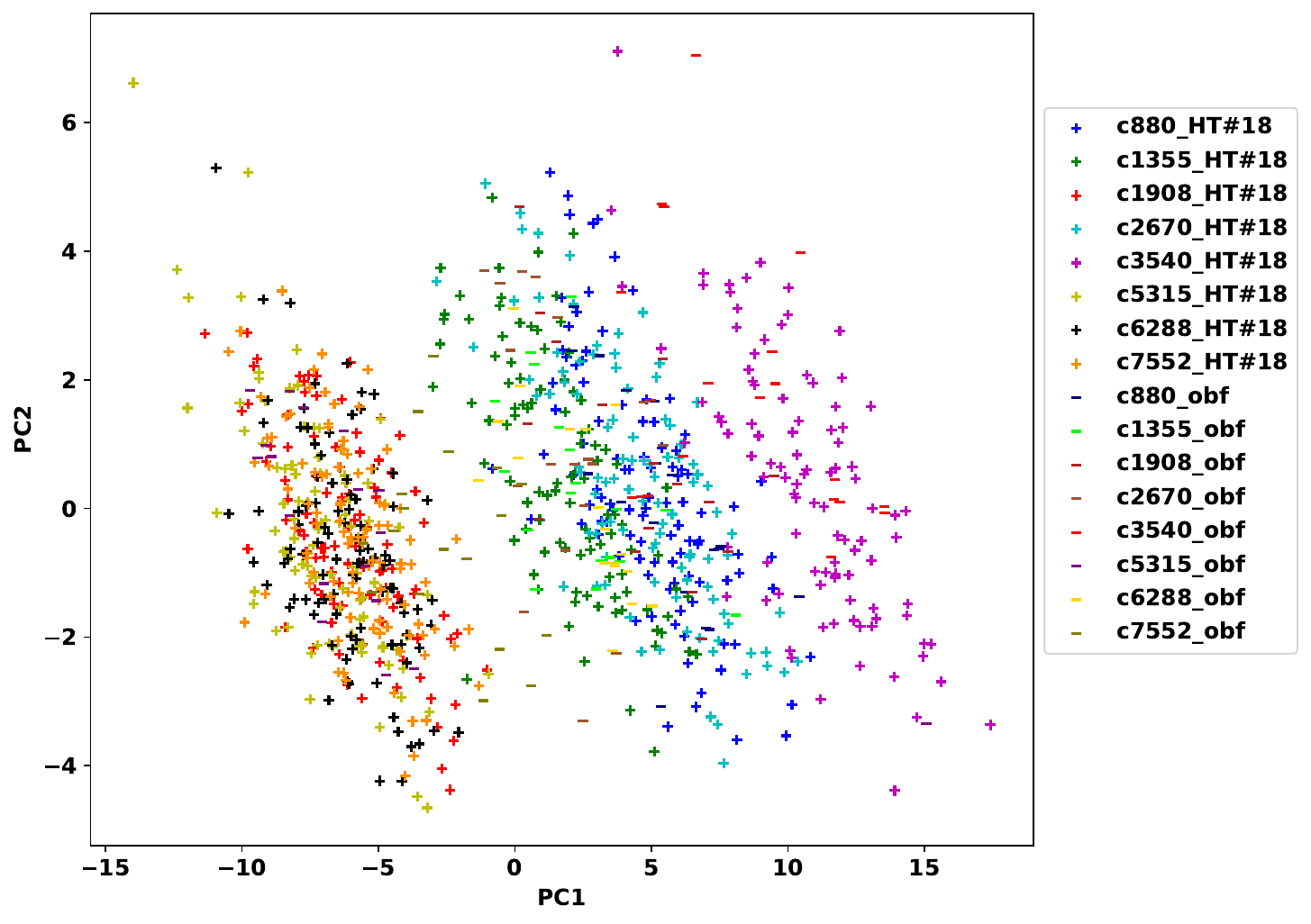}
\vspace{-5mm}
\caption{PCA analysis of hidden HTs vs.\ clean functionally transformed ISCAS-85 circuits}
\label{fig:ABC_HT_vs_ABC_clean}
\end{figure}

Figure~\ref{fig:ABC_HT_vs_ABC_clean} shows the outcome of the PCA analysis for the hidden versions of both infected (ending with \_HT\#18 suffix) and clean (ending with \_obf suffix) ISCAS-85 circuits. As can be seen, the HT-free instances are distributed among the HT-infected circuits, making it harder to find a boundary upon which they can be separated. It will be discussed in Section~\ref{subsec:HT_detection_analysis} that this will cause an issue for HW2VEC where it struggles to fully distinguish between the HT-free and HT-infected functionally transformed circuits. This figure emphasizes the need for a diversified dataset for training HT detectors. One solution could be to increase the number of ways circuits could be converted to different structures. 

\begin{figure}[!t]
\centering
\includegraphics[width=\linewidth]{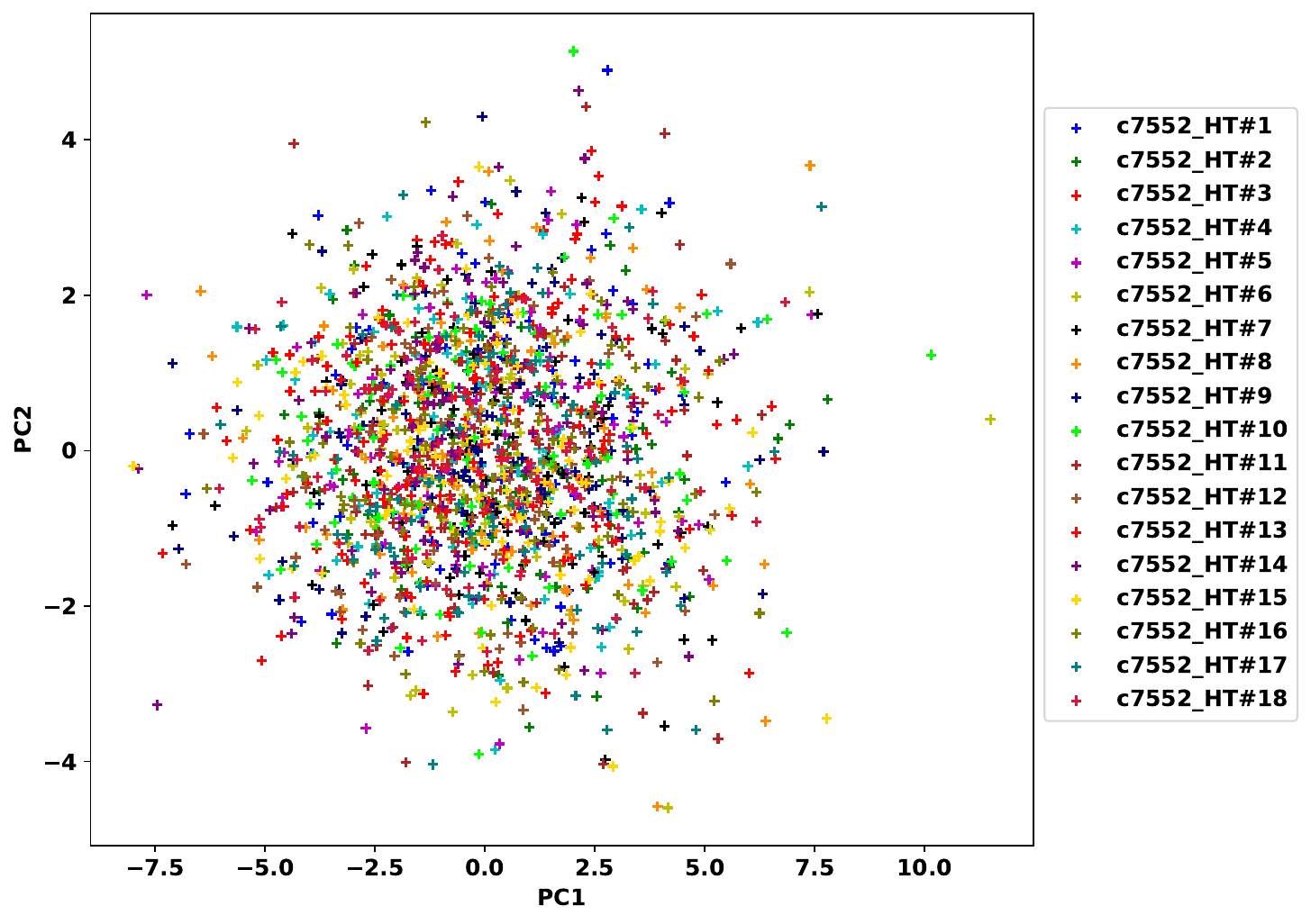}
\vspace{-5mm}
\caption{PCA analysis of all functional transformation techniques for $c7552$}
\label{fig:ABC_HT}
\vspace{-5mm}
\end{figure}

Figure~\ref{fig:ABC_HT} shows the PCA analysis for the 18 hidden versions of HT-infected $c7552$ circuits where no particular pattern for each \change{functional restructuring} approach can be derived. The reason can be sought in ABC’s AIG structures in its \change{functional restructuring} techniques. Accordingly, the PCA analysis will not show meaningful differences in each \change{functional restructuring method}; however, each generated benchmark provides more variability needed for training better HT detectors. As ABC is not designed to address the issues of hardware security, one future research direction for the research community might be investing time into building more \change{functionally equivalent} transformed circuits that may help diversify the HT pool.

\subsection{HT Detection Analysis}
\label{subsec:HT_detection_analysis}

\begin{figure*}[t!]
\centering
    
\begin{subfigure}[t]{0.44\textwidth}
    \centering
    \includegraphics[width=\linewidth]{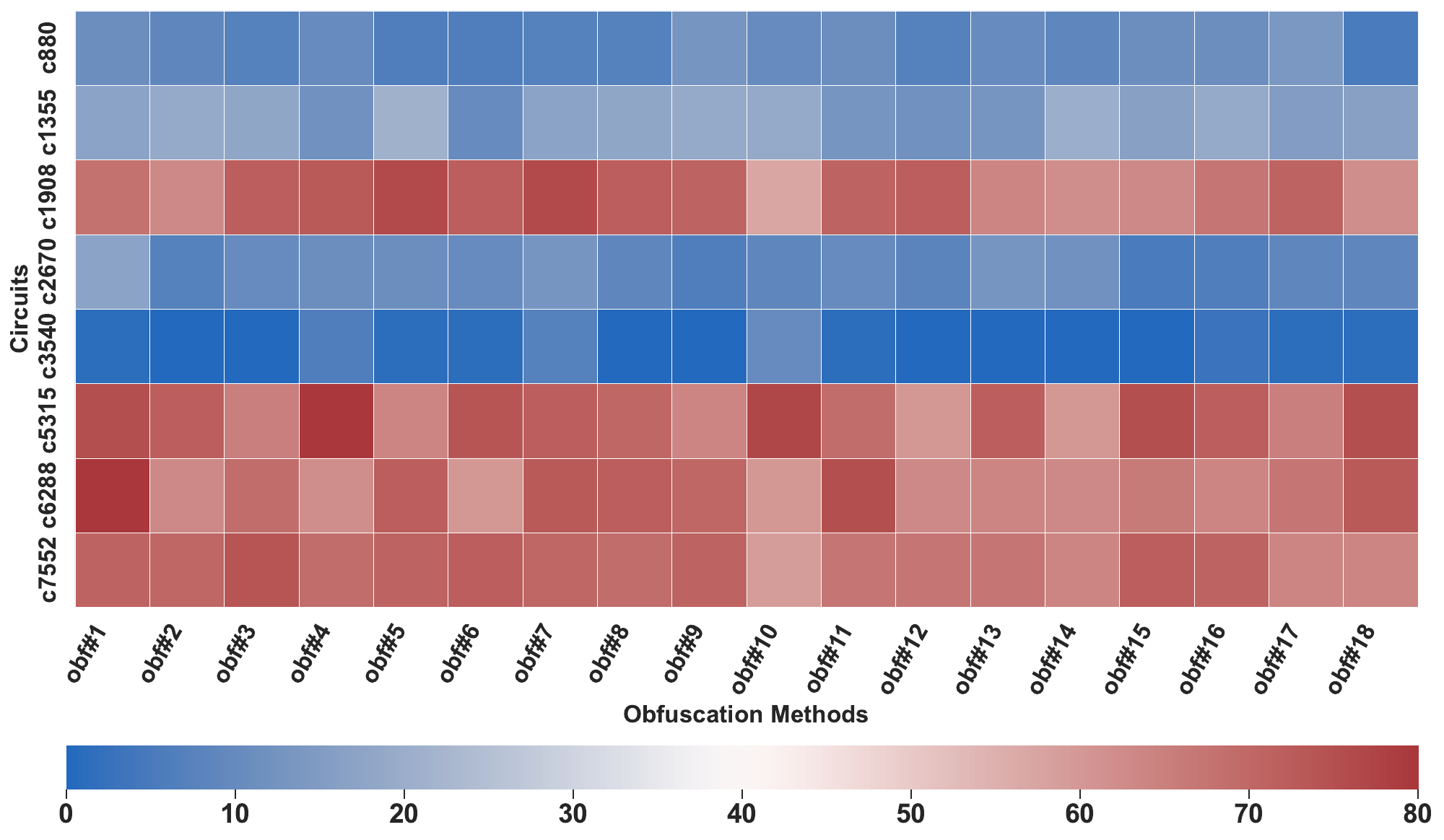}
    \caption{The HT detection accuracy spans between 0\% and 80\% for $S_{1}$}
    \label{fig:heatmap_TJ_RTL}
\end{subfigure}
\vspace{1mm}
\begin{subfigure}[t]{0.44\textwidth}
    \centering
    \includegraphics[width=\linewidth]{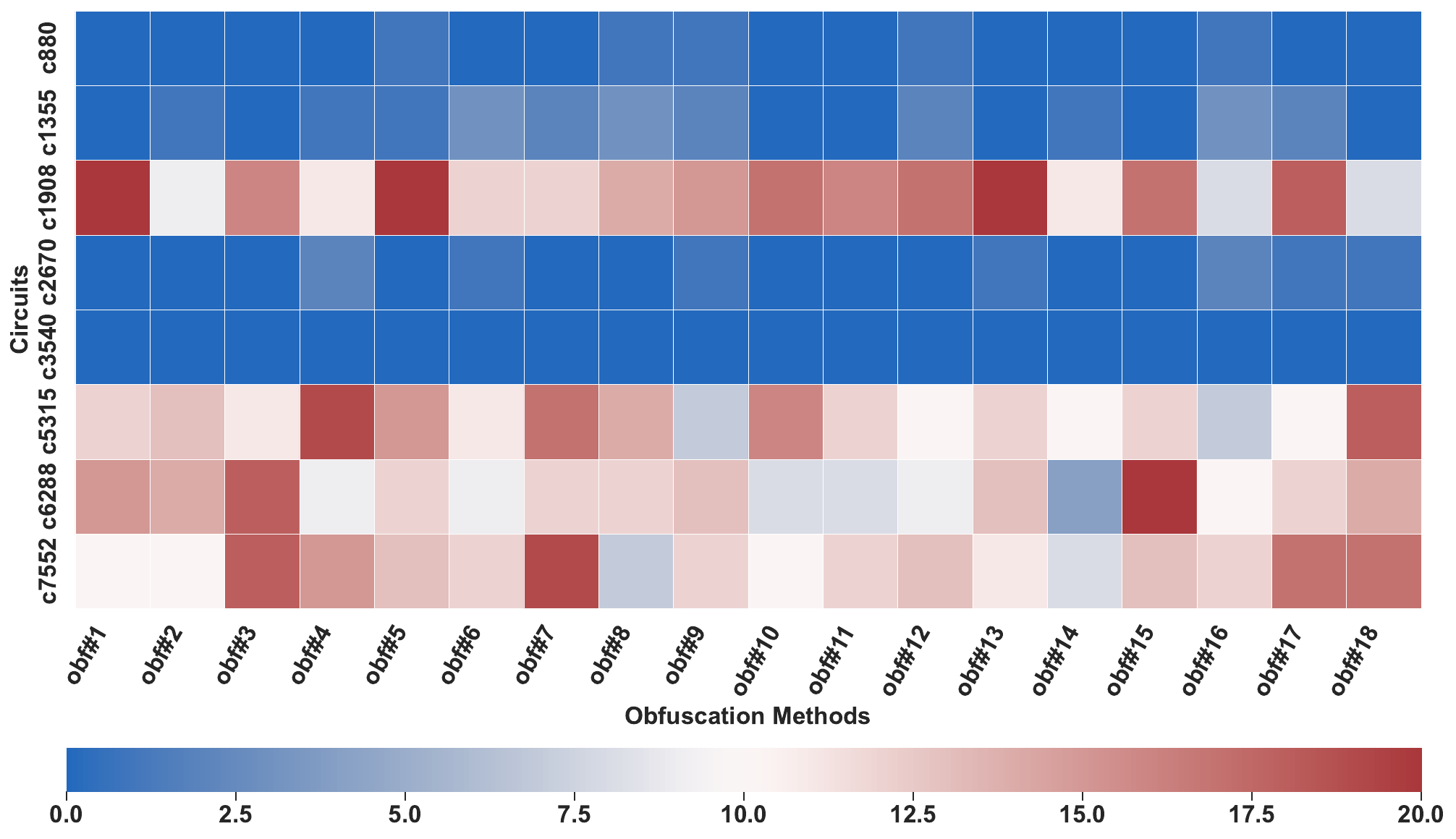}
    \caption{The HT detection accuracy spans between 0\% and 20\% for $S_{2}$}
    \label{fig:heatmap_TJ_RTL_ISCAS}
\end{subfigure}
\caption{HT detection accuracy of 18 functionally equivalent transformation methods trained with a) $S_{1}$ and b) $S_{2}$ for ISCAS-85 benchmarks. 
}
\label{fig:HT_detection_heatmap}
\vspace{-3mm}
\end{figure*}

Figure~\ref{fig:HT_detection_heatmap} shows the heatmap of HT detection accuracy percentages ($TPs$) for the benchmark circuits using HW2VEC trained with both $S_{1}$ and $S_{2}$. Each circuit-functional-restructuring method pair contains $100$ HT-infected circuits generated by the RL-inserter and \change{functionally equivalent} transformations using ABC. The detection accuracy in Figure~\ref{fig:heatmap_TJ_RTL} ranges between $0\%$ and $80\%$ for $S_{1}$ while the same figure ranges between $0\%$ and $20\%$ in Figure~\ref{fig:heatmap_TJ_RTL_ISCAS} for $S_{2}$. In both detection scenarios, the circuits are divided into two groups: 
\begin{enumerate}
\item $c880$, $c1355$, $c2670$, and $c3540$
\item  $c1908$, $c5315$, $c6288$, and $c7552$
\end{enumerate}
While HW2VEC detects up to $80\%$ of HTs in the second group under $S1$, it significantly underperforms with the first group. The situation worsens under $S2$, where the detector fails to classify HT-infected circuits in group `1` while the figures are only slightly better for group ‘2’. The underlying reason can be sought with the mixture of labels in each $S_{1}$ and $S_{2}$ and the unseen hidden data. The extra HT-free labels in $S_{2}$ have biased the detector to classify more instances as HT-free.



\begin{figure}[t!]
\centering

\begin{subfigure}[t]{0.42\textwidth}
    \centering
    \includegraphics[width=\linewidth]{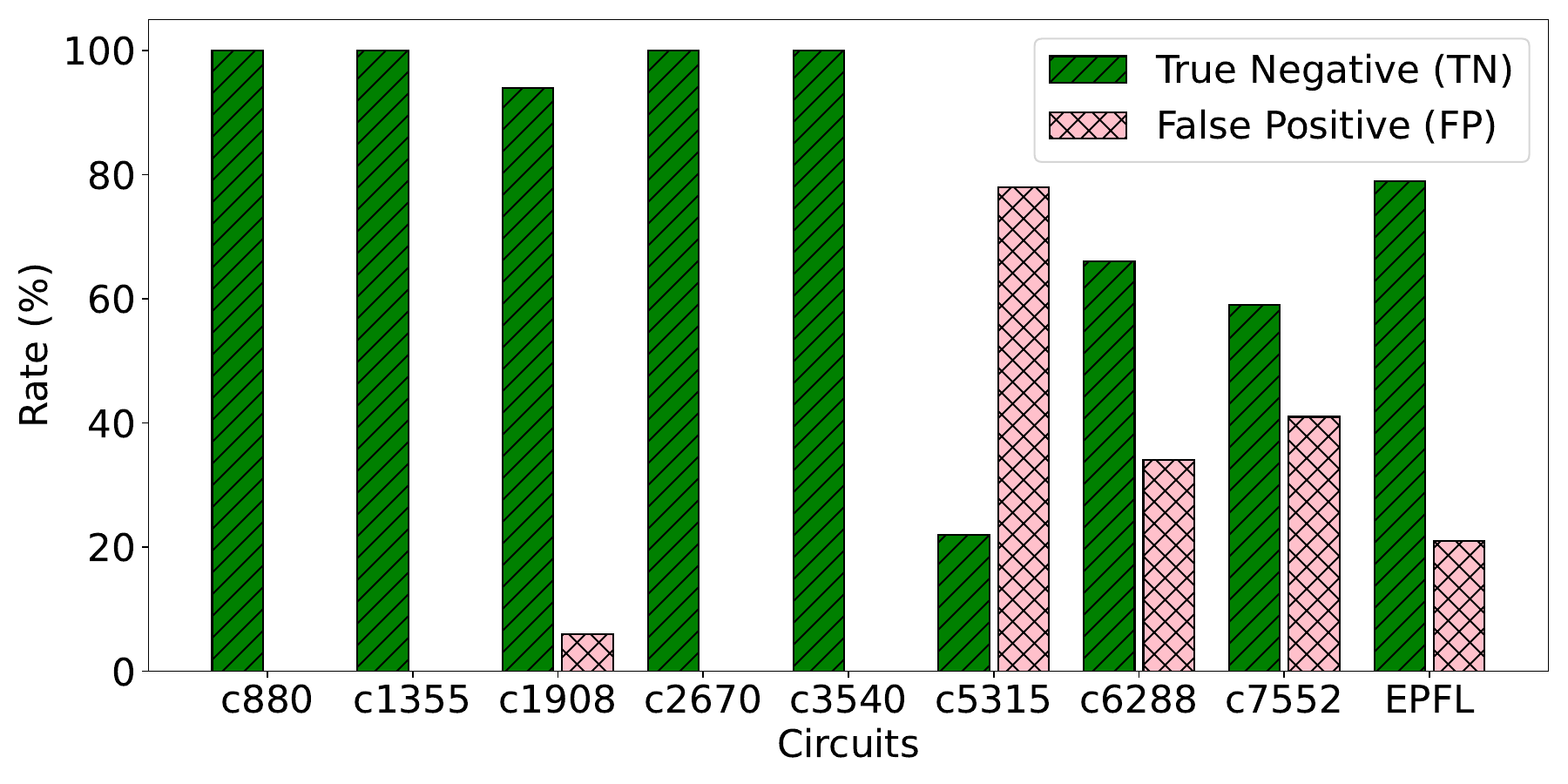}
    \vspace{-5mm}
    \caption{$TN$ and $FP$ percentages for $S_{1}$}
    \label{subfig:HW2VEC_TP_FN_S1}
\end{subfigure}

\vspace{1mm}

\begin{subfigure}[t]{0.42\textwidth}
    \centering
    \includegraphics[width=\linewidth]{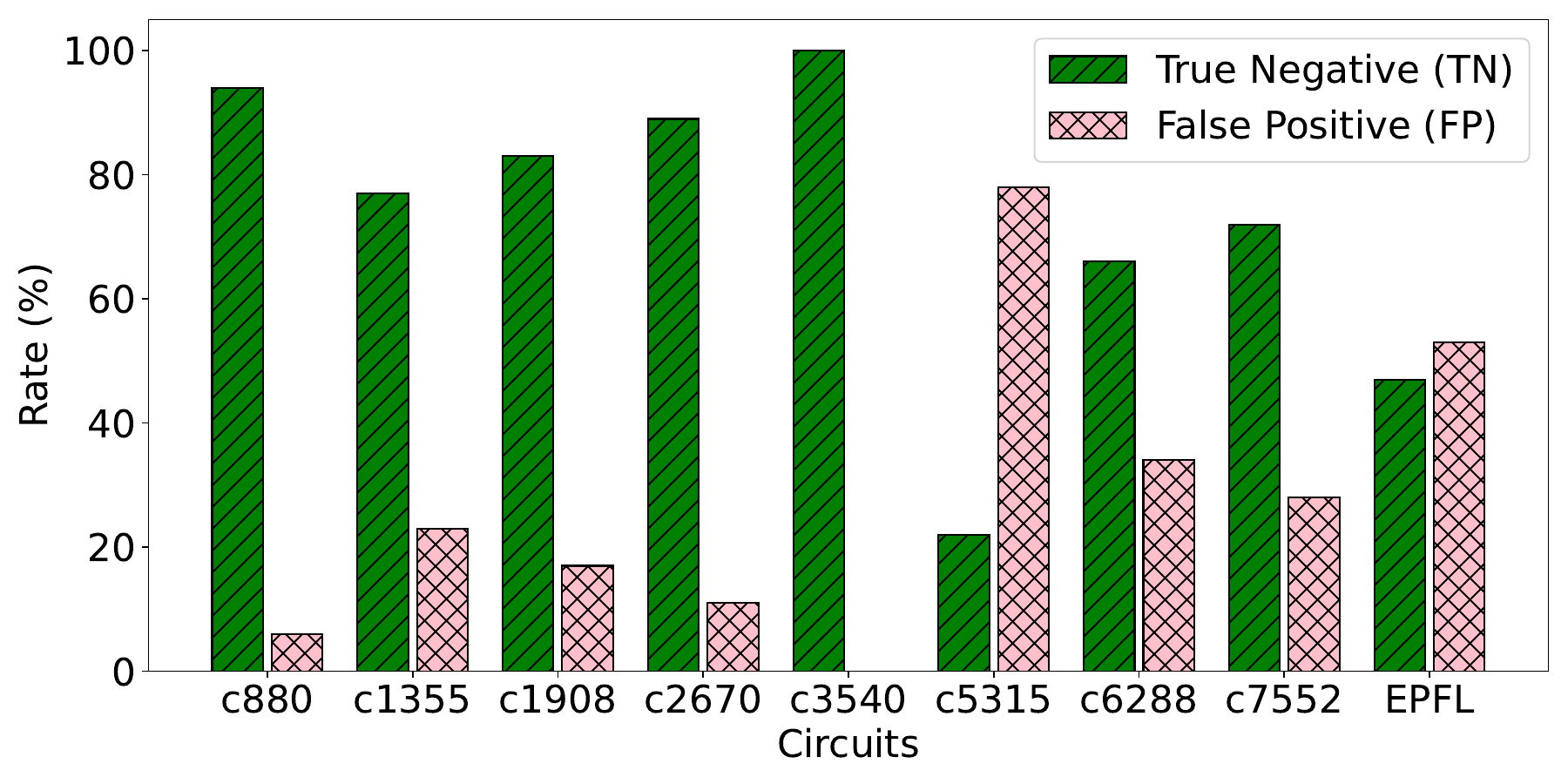}
    \caption{$TN$ and $FP$ percentages for $S_{2}$}
    \label{subfig:HW2VEC_TP_FN_S2}
\end{subfigure}
\caption{The comparison of $TN$ and $FP$ percentages of functionally equivalent transformed HT-free circuits for $S_{1}$ and $S_{2}$ detection scenarios
}
\label{fig:HW2VEC_TPs_FN}
\end{figure}

Figure~\ref{fig:HW2VEC_TPs_FN} shows the $TN$ and $FP$ rates for ISCAS-85 HT-free circuits trained with $S_{1}$ and $S_{2}$. Each circuit has $18$ variations of its original HT-free version. We also include $19$ combinational HT-free circuits from the EPFL synthesis benchmark~\cite{amaru2015epfl} to test the detection capabilities of HW2VEC further. By comparing figures~\ref{subfig:HW2VEC_TP_FN_S1} and~\ref{subfig:HW2VEC_TP_FN_S2}, we see that the $FP$ percentage is higher in the $S_{1}$ scenario. Again, due to the limited training data, HW2VEC is biased towards classifying most circuits as HT-free or HT-infected with a slight change in the mixture of training data labels. The comparison of $S_{1}$ and $S_{2}$ in figures~\ref{fig:HT_detection_heatmap} and~\ref{fig:HW2VEC_TPs_FN} shows that $FP$ rates are higher under $S_{1}$ and the detector is biased toward classifying more instances as HT-infected. The same conclusion can be made about $S_{2}$, where higher $TN$ rates lead to lower $FP$ rates. The circuits in the EPFL benchmark have substantial $FP$ rates in both $S_{1}$ and $S_{2}$.

\change{We also compare \textit{Seeker1} with two existing benchmarks introduced in~\cite{cruz2018automated,gohil2022attrition}. Both benchmarks only contain HT-infected circuits and HTs are inserted using low signal switching nets. For~\cite{cruz2018automated}, we train HW2VEC under $S_{1}$ and we report the detection accuracies for 4 reported ISCAS-85 benchmarks, \textit{c2670, c3540, c5315, c6288}. The detection figures are 100\%, 0\%, 70\%, and 0\%, respectively. In~\cite{gohil2022attrition}, there are two ISCAS-85 benchmarks: $c6288$ and $c7552$. The detection rates are 10\% and 90\%, respectively. Compared with Figure~\ref{fig:heatmap_TJ_RTL}, \textit{Seeker1} evades HT detection more consistently throughout the entire benchmark. It is important to note that the insertion criteria of~\cite{sarihi2023trojanframework,sarihi2022hardware} are inherently different from that of ~\cite{cruz2018automated,gohil2022attrition}.  In the future, we plan to investigate the impact of various HT insertion and functional restructuring strategies on HT detectors.}

\begin{figure}[!t]
\centering
\includegraphics[width=\linewidth]{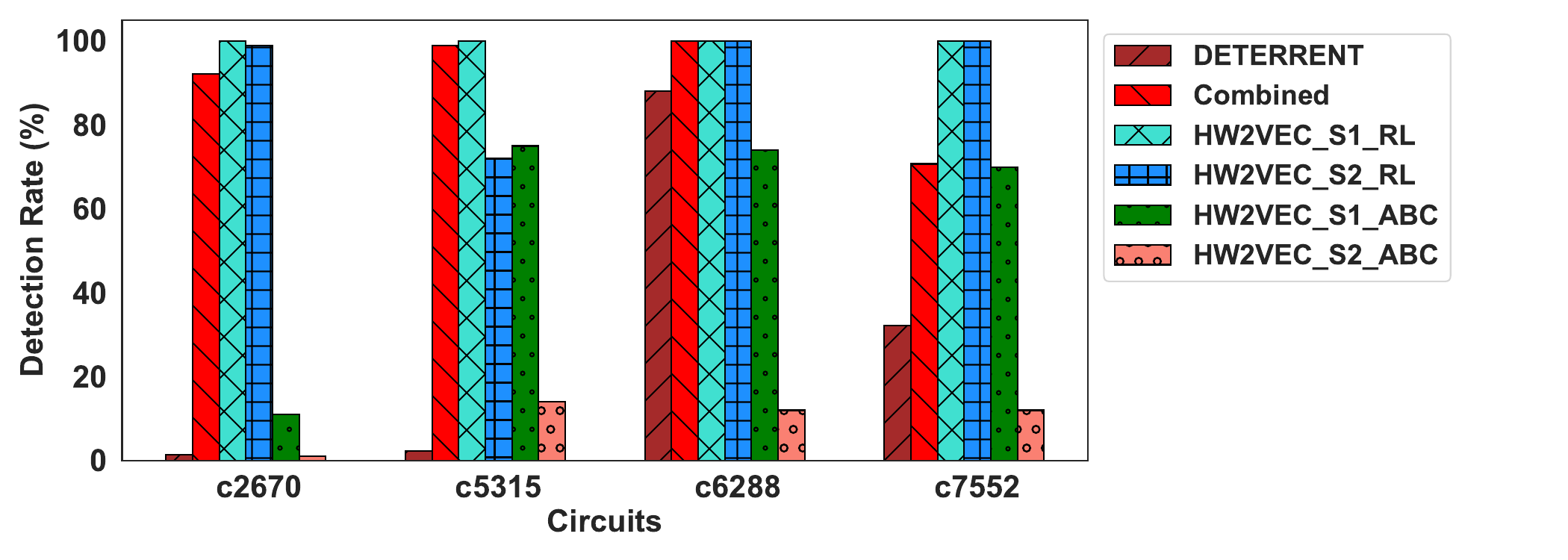}
\vspace{-5mm}
\caption{The detection rate (TPs) of DETERRENT~\cite{gohil2022deterrent}, RL\_HT\_DETECT~\cite{sarihi2023trojanframework}, and HW2VEC~\cite{yu2021hw2vec} for hidden HT-infected ISCAS-85 circuits}
\vspace{-5mm}
\label{fig:joint_detection}
\end{figure}

Figure~\ref{fig:joint_detection} shows the detection percentage for RL\_HT\_DETECT (\textit{Combined}), DETERRENT, and HW2VEC for the original HT-infected circuits (ending with \_RL suffix) and their ABC-\change{functionally equivalent transformed} versions (ending with \_ABC suffix). The $x$ axis shows the benchmark circuits and the $y$ axis shows the HT detection accuracy (\textit{TPs}) as a percentage. To fairly compare against DETERRENT, we only mention the four ISCAS-85 benchmarks studied in the DETERRENT paper. 
As can be seen, the detection accuracy of RL\_HT\_DETECT is higher than DETERRENT in all four circuits; however, the difference is more substantial in $c2670$ and $c5315$. The reason can be sought into  multi-criteria~\cite{sarihi2023multi} versus single criterion~\cite{gohil2022deterrent} HT detection. As for HW2VEC's detection rate under $S_{1}$ and $S_{2}$ for the baseline RL benchmarks is nearly $100\%$. The situation differs for the \change{functionally equivalent transformed} HTs, with lower HT detection rates. We plug the $FP$ and $FN$ values in Equation~\ref{equ:metric}, assuming a threat model with $\alpha = 10$. Accordingly, the average $Conf. Val$ for RL\_HT\_DETECT, DETERRENT, HW2VEC\_$S_{1}$, and HW2VEC\_$S_{2}$ detectors over all circuits are 5.55, 1.31, 3.33, and 1.66, respectively. Thus, the security engineer can invest more trust in RL\_HT\_DETECT with the assumed $\alpha$.
 
\section{Discussion}
\label{sec:discussion}

The Seeker’s Dilemma for HT Detection comparison provides a promising approach to improve HT detection since the approach and methodology capture the reality of HT detection. Our analysis provides insights into how good \textit{Seeker1} as a benchmark is, and we believe that a benchmark like this should exist for three years from the publication of this work. The introduced generalized approach and innovations in HT insertion and detection provide a model that the research community, as a whole, should greatly benefit from. This brief discussion highlights the shortcomings of our approach and what needs to be addressed next. \change{The scalability of our approach is closely related to the computation capabilities of the attacker; the netlist restructuring process can be pushed as far as the ABC restructuring capabilities. }

We have not tested our \textit{Seeker1} benchmark with test-driven HT detectors such as those that use side-channels measurements. We hypothesize that HT side-channel detection methods will not easily classify Seeker’s Dilemma benchmarks since those approaches compare circuit characteristics for a set of manufactured ICs, looking for outliers in the set that would suggest modification. Our functional restructuring techniques in our benchmark generation make this approach very difficult. This, however, is an untested hypothesis since open-source side-channel tools do not exist.

One major shortcoming of the \textit{Seeker1} is that we need to move HT insertion and detection closer to commercial designs. This means that both sequential circuits and the benchmark size need to be updated. From a complexity theory perspective, the Seeker’s Dilemma problem is a computationally very hard problem as these two characteristics of a circuit are included. The size of a circuit makes the \textit{Hide\&Seek} search space larger, and sequential circuits make both the insertion of HTs and the detection harder. \change{Finally, from the practicality standpoint, our proposed flow requires access to the design netlist or source code in the pre-silicon stage. While third-party IP providers might encrypt source code and not be readily available, aligned with our threat model, we assume that the post-synthesis netlist is available for detectors such as HW2VEC.}

\section{Conclusion}
\label{C}
\change{This work provides a realistic definition for the state-of-the-art problem of HT detection by describing the problem as an extension of Hide\&Seek on a graph, i.e., The Seeker’s Dilemma.  Our problem statement is based on the fact that an HT detection agent is unaware of whether circuits are infected by HTs or not. We used the Seeker’s Dilemma paradigm to propose a new HT benchmarking strategy for a more accurate evaluation of detection methods. We anticipate that this innovative problem statement and the benchmarking strategy will help the community explore innovative ideas for HT detection by facilitating accurate quality assessments of existing detection methods. Finally, we created a combinational benchmark following a prescribed method and have released this benchmark to the general community to steer work in HT detection and insertion on a new path. We show how existing HT detection methods perform on this benchmark, making this the first attempt to identify HTs in the released benchmark.}




\balance
\bibliographystyle{References-Style/IEEEtran}
\bibliography{References-Style/IEEEabrv, References-Style/IEEEexample}

\begin{thebibliography}{10}
\providecommand{\url}[1]{#1}
\csname url@samestyle\endcsname
\providecommand{\newblock}{\relax}
\providecommand{\bibinfo}[2]{#2}
\providecommand{\BIBentrySTDinterwordspacing}{\spaceskip=0pt\relax}
\providecommand{\BIBentryALTinterwordstretchfactor}{4}
\providecommand{\BIBentryALTinterwordspacing}{\spaceskip=\fontdimen2\font plus
\BIBentryALTinterwordstretchfactor\fontdimen3\font minus
  \fontdimen4\font\relax}
\providecommand{\BIBforeignlanguage}[2]{{%
\expandafter\ifx\csname l@#1\endcsname\relax
\typeout{** WARNING: IEEEtran.bst: No hyphenation pattern has been}%
\typeout{** loaded for the language `#1'. Using the pattern for}%
\typeout{** the default language instead.}%
\else
\language=\csname l@#1\endcsname
\fi
#2}}
\providecommand{\BIBdecl}{\relax}
\BIBdecl

\bibitem{yu2021hw2vec}
S.-Y. Yu, R.~Yasaei, Q.~Zhou, T.~Nguyen, and M.~A. Al~Faruque, ``Hw2vec: A
  graph learning tool for automating hardware security,'' in \emph{2021 IEEE
  International Symposium on Hardware Oriented Security and Trust
  (HOST)}.\hskip 1em plus 0.5em minus 0.4em\relax IEEE, 2021, pp. 13--23.

\bibitem{shakya2017benchmarking}
B.~Shakya, T.~He, H.~Salmani, D.~Forte, S.~Bhunia, and M.~Tehranipoor,
  ``Benchmarking of hardware trojans and maliciously affected circuits,''
  \emph{Journal of Hardware and Systems Security}, vol.~1, no.~1, pp. 85--102,
  2017.

\bibitem{jain2021survey}
A.~Jain, Z.~Zhou, and U.~Guin, ``Survey of recent developments for hardware
  trojan detection,'' in \emph{2021 IEEE International Symposium on Circuits
  and Systems (ISCAS)}.\hskip 1em plus 0.5em minus 0.4em\relax IEEE, 2021, pp.
  1--5.

\bibitem{chakraborty2009mero}
R.~S. Chakraborty, F.~Wolff, S.~Paul, C.~Papachristou, and S.~Bhunia, ``Mero: A
  statistical approach for hardware trojan detection,'' in \emph{International
  Workshop on Cryptographic Hardware and Embedded Systems}.\hskip 1em plus
  0.5em minus 0.4em\relax Springer, 2009, pp. 396--410.

\bibitem{hasegawa2017trojan}
K.~Hasegawa, M.~Yanagisawa, and N.~Togawa, ``Trojan-feature extraction at
  gate-level netlists and its application to hardware-trojan detection using
  random forest classifier,'' in \emph{2017 IEEE International Symposium on
  Circuits and Systems (ISCAS)}.\hskip 1em plus 0.5em minus 0.4em\relax IEEE,
  2017, pp. 1--4.

\bibitem{pan2021automated}
Z.~Pan and P.~Mishra, ``Automated test generation for hardware trojan detection
  using reinforcement learning,'' in \emph{Proceedings of the 26th Asia and
  South Pacific Design Automation Conference}, 2021, pp. 408--413.

\bibitem{gohil2022deterrent}
V.~Gohil, S.~Patnaik, H.~Guo, D.~Kalathil, and J.~Rajendran, ``Deterrent:
  detecting trojans using reinforcement learning,'' in \emph{Proceedings of the
  59th ACM/IEEE Design Automation Conference}, 2022, pp. 697--702.

\bibitem{sarihi2023multi}
A.~Sarihi, P.~Jamieson, A.~Patooghy, and A.-H.~A. Badawy, ``Multi-criteria
  hardware trojan detection: A reinforcement learning approach,'' in \emph{2023
  IEEE 66th International Midwest Symposium on Circuits and Systems (MWSCAS)},
  2023, pp. 1093--1097.

\bibitem{trusthub}
``{Trust-Hub},'' \url{https://trust-hub.org/#/home}, accessed: 2022-12-19.

\bibitem{cruz2018automated}
J.~Cruz, Y.~Huang, P.~Mishra, and S.~Bhunia, ``An automated configurable trojan
  insertion framework for dynamic trust benchmarks,'' in \emph{2018 Design,
  Automation \& Test in Europe Conference \& Exhibition (DATE)}.\hskip 1em plus
  0.5em minus 0.4em\relax IEEE, 2018, pp. 1598--1603.

\bibitem{gohil2022attrition}
V.~Gohil, H.~Guo, S.~Patnaik, and J.~Rajendran, ``Attrition: Attacking static
  hardware trojan detection techniques using reinforcement learning,'' in
  \emph{Proceedings of the 2022 ACM SIGSAC Conference on Computer and
  Communications Security}, 2022, pp. 1275--1289.

\bibitem{sarihi2023trojanframework}
A.~Sarihi, A.~Patooghy, P.~Jamieson, and A.-H.~A. Badawy, ``Trojan playground:
  A reinforcement learning framework for hardware trojan insertion and
  detection,'' \emph{arXiv preprint arXiv:2305.09592}, 2023.

\bibitem{githubGitHubNMSUPEARLSeekersDilemmaHardwareTrojanBenchmarks}
``Seeker's dilemma hardware trojan-benchmarks: Functionally restructured
  hardware trojan benchmarks,''
  \url{https://github.com/NMSU-PEARL/Seeker-s-Dilemma-Hardware-Trojan-Benchmarks},
  [Accessed 27-02-2024].

\bibitem{xue2020ten}
M.~Xue, C.~Gu, W.~Liu, S.~Yu, and M.~O'Neill, ``Ten years of hardware trojans:
  a survey from the attacker's perspective,'' \emph{IET Computers \& Digital
  Techniques}, vol.~14, no.~6, pp. 231--246, 2020.

\bibitem{li2016survey}
H.~Li, Q.~Liu, and J.~Zhang, ``A survey of hardware trojan threat and
  defense,'' \emph{Integration}, vol.~55, pp. 426--437, 2016.

\bibitem{hoque2020hardware}
T.~Hoque, R.~S. Chakraborty, and S.~Bhunia, ``Hardware obfuscation and logic
  locking: A tutorial introduction,'' \emph{IEEE Design \& Test}, vol.~37,
  no.~3, pp. 59--77, 2020.

\bibitem{lyu2020scalable}
Y.~Lyu and P.~Mishra, ``Scalable activation of rare triggers in hardware
  trojans by repeated maximal clique sampling,'' \emph{IEEE Transactions on
  Computer-Aided Design of Integrated Circuits and Systems}, vol.~40, no.~7,
  pp. 1287--1300, 2020.

\bibitem{salmani2016cotdpaper}
H.~Salmani, ``Cotd: Reference-free hardware trojan detection and recovery based
  on controllability and observability in gate-level netlist,'' \emph{IEEE
  Transactions on Information Forensics and Security}, vol.~12, no.~2, pp.
  338--350, 2016.

\bibitem{gubbi2023hardware}
K.~I. Gubbi, B.~Saber~Latibari, A.~Srikanth, T.~Sheaves, S.~A.
  Beheshti-Shirazi, S.~M. PD, S.~Rafatirad, A.~Sasan, H.~Homayoun, and
  S.~Salehi, ``Hardware trojan detection using machine learning: A tutorial,''
  \emph{ACM Transactions on Embedded Computing Systems}, vol.~22, no.~3, pp.
  1--26, 2023.

\bibitem{goldstein1980scoap}
L.~H. Goldstein and E.~L. Thigpen, ``Scoap: Sandia
  controllability/observability analysis program,'' in \emph{Proceedings of the
  17th Design Automation Conference}, 1980, pp. 190--196.

\bibitem{salmani2013design}
H.~Salmani, M.~Tehranipoor, and R.~Karri, ``On design vulnerability analysis
  and trust benchmarks development,'' in \emph{2013 IEEE 31st international
  conference on computer design (ICCD)}.\hskip 1em plus 0.5em minus 0.4em\relax
  IEEE, 2013, pp. 471--474.

\bibitem{sarihi2022hardware}
A.~Sarihi, A.~Patooghy, P.~Jamieson, and B.~A.-H. A., ``Hardware trojan
  insertion using reinforcement learning,'' in \emph{Proceedings of the Great
  Lakes Symposium on VLSI 2022}, 2022, pp. 139--142.

\bibitem{liakos2022gainesis}
K.~G. Liakos, G.~K. Georgakilas, F.~C. Plessas, and P.~Kitsos, ``Gainesis:
  Generative artificial intelligence netlists synthesis,'' \emph{Electronics},
  vol.~11, no.~2, p. 245, 2022.

\bibitem{yu2019improved}
S.~Yu, W.~Liu, and M.~O'Neill, ``An improved automatic hardware trojan
  generation platform,'' in \emph{2019 IEEE Computer Society Annual Symposium
  on VLSI (ISVLSI)}.\hskip 1em plus 0.5em minus 0.4em\relax IEEE, 2019, pp.
  302--307.

\bibitem{kamhoua2016game}
C.~A. Kamhoua, H.~Zhao, M.~Rodriguez, and K.~A. Kwiat, ``A game-theoretic
  approach for testing for hardware trojans,'' \emph{IEEE Transactions on
  Multi-Scale Computing Systems}, vol.~2, no.~3, pp. 199--210, 2016.

\bibitem{saad2017hardware}
W.~Saad, A.~Sanjab, Y.~Wang, C.~A. Kamhoua, and K.~A. Kwiat, ``Hardware trojan
  detection game: A prospect-theoretic approach,'' \emph{IEEE Transactions on
  Vehicular Technology}, vol.~66, no.~9, pp. 7697--7710, 2017.

\bibitem{kahneman2013prospect}
D.~Kahneman and A.~Tversky, ``Prospect theory: An analysis of decision under
  risk,'' in \emph{Handbook of the fundamentals of financial decision making:
  Part I}.\hskip 1em plus 0.5em minus 0.4em\relax World Scientific, 2013, pp.
  99--127.

\bibitem{das2020think}
T.~Das, A.~R. Eldosouky, and S.~Sengupta, ``Think smart, play dumb: Analyzing
  deception in hardware trojan detection using game theory,'' in \emph{2020
  International Conference on Cyber Security and Protection of Digital Services
  (Cyber Security)}.\hskip 1em plus 0.5em minus 0.4em\relax IEEE, 2020, pp.
  1--8.

\bibitem{brahma2021game}
S.~Brahma, L.~Njilla, and S.~Nan, ``Game theoretic hardware trojan testing
  under cost considerations,'' in \emph{International Conference on Decision
  and Game Theory for Security}.\hskip 1em plus 0.5em minus 0.4em\relax
  Springer, 2021, pp. 251--270.

\bibitem{nan2023game}
S.~Nan, L.~Njilla, S.~Brahma, and C.~A. Kamhoua, ``Game and prospect theoretic
  hardware trojan testing,'' in \emph{2023 57th Annual Conference on
  Information Sciences and Systems (CISS)}.\hskip 1em plus 0.5em minus
  0.4em\relax IEEE, 2023, pp. 1--6.

\bibitem{gohil2021games}
V.~Gohil, M.~Tressler, K.~Sipple, S.~Patnaik, and J.~Rajendran, ``Games,
  dollars, splits: A game-theoretic analysis of split manufacturing,''
  \emph{IEEE Transactions on Information Forensics and Security}, vol.~16, pp.
  5077--5092, 2021.

\bibitem{alpern2006theory}
S.~Alpern and S.~Gal, \emph{The theory of search games and rendezvous}.\hskip
  1em plus 0.5em minus 0.4em\relax Springer Science \& Business Media, 2006,
  vol.~55.

\bibitem{isaacs1965differential}
R.~Isaacs, ``Differential games, siam series in applied mathematics,'' 1965.

\bibitem{stone1976theory}
L.~D. Stone, \emph{Theory of optimal search}.\hskip 1em plus 0.5em minus
  0.4em\relax Elsevier, 1976.

\bibitem{chapman2014playing}
M.~Chapman, G.~Tyson, P.~McBurney, M.~Luck, and S.~Parsons, ``Playing
  hide-and-seek: an abstract game for cyber security,'' in \emph{Proceedings of
  the 1st International Workshop on Agents and CyberSecurity}, 2014, pp. 1--8.

\bibitem{jamieson2018benchmarking}
P.~Jamieson, A.~Sanaullah, and M.~Herbordt, ``Benchmarking heterogeneous hpc
  systems including reconfigurable fabrics: Community aspirations for ideal
  comparisons,'' in \emph{2018 IEEE High Performance extreme Computing
  Conference (HPEC)}.\hskip 1em plus 0.5em minus 0.4em\relax IEEE, 2018, pp.
  1--6.

\bibitem{jamieson2010benchmarking}
P.~Jamieson, T.~Becker, P.~Y. Cheung, W.~Luk, T.~Rissa, and T.~Pitk{\"a}nen,
  ``Benchmarking and evaluating reconfigurable architectures targeting the
  mobile domain,'' \emph{ACM Transactions on Design Automation of Electronic
  Systems (TODAES)}, vol.~15, no.~2, pp. 1--24, 2010.

\bibitem{bryan1985iscas}
D.~Bryan, ``The iscas'85 benchmark circuits and netlist format,'' \emph{North
  Carolina State University}, vol.~25, p.~39, 1985.

\bibitem{reshma2019hardware}
K.~Reshma, M.~Priyatharishini, and M.~Nirmala~Devi, ``Hardware trojan detection
  using deep learning technique,'' in \emph{Soft Computing and Signal
  Processing: Proceedings of ICSCSP 2018, Volume 2}.\hskip 1em plus 0.5em minus
  0.4em\relax Springer, 2019, pp. 671--680.

\bibitem{jyothi2017taint}
V.~Jyothi, P.~Krishnamurthy, F.~Khorrami, and R.~Karri, ``Taint: Tool for
  automated insertion of trojans,'' in \emph{2017 IEEE International Conference
  on Computer Design (ICCD)}.\hskip 1em plus 0.5em minus 0.4em\relax IEEE,
  2017, pp. 545--548.

\bibitem{pan2002teaching}
Y.~Pan, ``Teaching parallel programming using both high-level and low-level
  languages,'' \emph{Computational Science—ICCS 2002}, pp. 888--897, 2002.

\bibitem{mishchenko2018integrating}
A.~Mishchenko and R.~Brayton, ``Integrating an aig package, simulator, and sat
  solver,'' in \emph{International Workshop on Logic and Synthesis (IWLS)},
  2018, pp. 11--16.

\bibitem{chowdhury2021openabc}
A.~B. Chowdhury, B.~Tan, R.~Karri, and S.~Garg, ``Openabc-d: A large-scale
  dataset for machine learning guided integrated circuit synthesis,''
  \emph{arXiv preprint arXiv:2110.11292}, 2021.

\bibitem{mishchenko2005fraigs}
A.~Mishchenko, S.~Chatterjee, R.~Jiang, and R.~K. Brayton, ``Fraigs: A unifying
  representation for logic synthesis and verification,'' ERL Technical Report,
  Tech. Rep., 2005.

\bibitem{brayton2010abc}
R.~Brayton and A.~Mishchenko, ``Abc: An academic industrial-strength
  verification tool,'' in \emph{Computer Aided Verification: 22nd International
  Conference, CAV 2010, Edinburgh, UK, July 15-19, 2010. Proceedings 22}.\hskip
  1em plus 0.5em minus 0.4em\relax Springer, 2010, pp. 24--40.

\bibitem{mishcenko}
A.~Mishchenko, ``Introduction to logic synthesis with abc,''
  \url{http://cc.ee.ntu.edu.tw/~jhjiang/instruction/courses/fall14-lsv/lec01-abc_4p.pdf}.

\bibitem{hansen1999unveiling}
M.~C. Hansen, H.~Yalcin, and J.~P. Hayes, ``Unveiling the iscas-85 benchmarks:
  A case study in reverse engineering,'' \emph{IEEE Design \& Test of
  Computers}, vol.~16, no.~3, pp. 72--80, 1999.

\bibitem{chollet2021deep}
F.~Chollet, \emph{Deep learning with Python}.\hskip 1em plus 0.5em minus
  0.4em\relax Simon and Schuster, 2021.

\bibitem{bro2014principal}
R.~Bro and A.~K. Smilde, ``Principal component analysis,'' \emph{Analytical
  methods}, vol.~6, no.~9, pp. 2812--2831, 2014.

\bibitem{liu2022pca}
P.~Liu, L.~Wu, Z.~Zhang, D.~Xiao, X.~Zhang, and L.~Wang, ``A pca based svm
  hardware trojan detection approach,'' in \emph{2022 IEEE 16th International
  Conference on Anti-counterfeiting, Security, and Identification
  (ASID)}.\hskip 1em plus 0.5em minus 0.4em\relax IEEE, 2022, pp. 1--5.

\bibitem{shende2016side}
R.~Shende and D.~D. Ambawade, ``A side channel based power analysis technique
  for hardware trojan detection using statistical learning approach,'' in
  \emph{2016 thirteenth international conference on wireless and optical
  communications networks (WOCN)}.\hskip 1em plus 0.5em minus 0.4em\relax IEEE,
  2016, pp. 1--4.

\bibitem{wang2013malicious}
L.~Wang, H.~Xie, and H.~Luo, ``Malicious circuitry detection using transient
  power analysis for ic security,'' in \emph{2013 International Conference on
  Quality, Reliability, Risk, Maintenance, and Safety Engineering
  (QR2MSE)}.\hskip 1em plus 0.5em minus 0.4em\relax IEEE, 2013, pp. 1164--1167.

\bibitem{amaru2015epfl}
L.~Amar{\'u}, P.-E. Gaillardon, and G.~De~Micheli, ``The epfl combinational
  benchmark suite,'' in \emph{Proceedings of the 24th International Workshop on
  Logic \& Synthesis (IWLS)}, no. CONF, 2015.

\end{thebibliography}

\end{document}